\documentclass[11pt,fleqn,a4paper,superscriptaddress]{revtex4}

\usepackage{graphicx}
\usepackage{dcolumn}
\usepackage{amsmath}
\usepackage{color}
\begin{document}

\title{Charge disproportionation and site-selective local magnetic moments in the post-perovskite-type Fe$_2$O$_3$ under ultra-high pressures}

\author{Ivan Leonov}
\affiliation{M.N. Miheev Institute of Metal Physics, Russian Academy of Sciences, 620137 Yekaterinburg, Russia}
\affiliation{Materials Modeling and Development Laboratory, National University of Science and Technology 'MISIS', 119049 Moscow, Russia}

\author{Gregory Kh. Rozenberg}
\affiliation{School of Physics and Astronomy, Tel Aviv University, 69978, Tel Aviv, Israel}

\author{Igor A. Abrikosov}
\affiliation{Department of Physics, Chemistry and Biology (IFM), Link\"oping University, SE-58183 Link\"oping, Sweden}
\affiliation{Materials Modeling and Development Laboratory, National University of Science and Technology 'MISIS', 119049 Moscow, Russia}

\begin{abstract}
The archetypal $3d$ Mott insulator hematite, Fe$_2$O$_3$, is one of the basic oxide components playing an important role in mineralogy of Earth's lower mantle. Its high pressure-temperature behavior, such as the electronic properties, equation of state, and phase stability is of fundamental importance for understanding the properties and evolution of the Earth's interior.
Here, we study the electronic structure, magnetic state, and lattice stability of Fe$_2$O$_3$ at ultra-high pressures using 
the density functional plus dynamical mean-field theory (DFT+DMFT) approach. In the vicinity of a Mott transition, Fe$_2$O$_3$ is found to exhibit a series of complex electronic, magnetic, and structural transformations. 
%
%
In particular, it makes a phase transition to a metal with a post-perovskite crystal structure and site-selective local moments upon compression above 75 GPa.
We show that the site-selective phase transition is accompanied by a charge disproportionation of Fe ions, with Fe$^{3\pm \delta}$ and $\delta \sim 0.05$-$0.09$, implying a complex interplay between electronic correlations and the lattice.
Our results suggest that site-selective local moments in Fe$_2$O$_3$ persist up to ultra-high pressures of $\sim$200-250 GPa, i.e., sufficiently above the core-mantle boundary. The latter can have important consequences for understanding of the velocity and density anomalies in the Earth's lower mantle.
\end{abstract}


\maketitle

E-mail: ivan.v.leonov@yandex.ru (I.L.) and igor.abrikosov@liu.se (I.A.A.).


\pagebreak

\section*{Introduction}

Being model objects for studying the Mott transition phenomenon, the iron-bearing oxides play an important role in the mineralogy of Earth's lower mantle and outer core \cite{Nature.436.377,GRL.38.L23308,PhysRevLett.110.228501,RevGeophys.51.244,Buffett2007,PNAS.106.5508,NatCommun.7.10661,Nature.570.102}. Because of their complex electronic, magnetic, and crystal structure behavior under high pressure-temperature conditions, these compounds have been of considerable recent interest  \cite{RevModPhys.70.1039,PhysRevLett.82.4663,PhysRevLett.102.146402,PNAS.106.5508,NatCommun.7.10661,Nature.570.102,PhysRevX.8.031059}. It is known that upon compression these materials exhibit a magnetic collapse -- a crossover from a high-spin to low-spin state of iron ions, resulting in drastic changes of their physical properties \cite{Science.275.654,PhysRevB.92.085142,PhysRevB.94.155135,PhysRevB.96.075136}. In fact, the anomalous behavior of their bulk modulus, density, and elastic properties is essential to understanding the seismic observations and dynamic processes in the Earth's lower mantle and outer core \cite{Nature.436.377,GRL.38.L23308,PhysRevLett.110.228501,RevGeophys.51.244,Buffett2007}, e.g., for interpretation of the anomalous seismic behavior at the bottom 400 km of Earth's mantle, in the so-called D'' region.
The high-pressure electronic and structural properties of ferric oxide, hematite ($\alpha$-Fe$_2$O$_3$), the ``classical'' example of a Mott insulating material, is of particular interest for both fundamental science and technological applications. Its high-pressure properties, such as, rich allotropic behavior, release of oxygen resulting in the appearance of a homologous series of $n$FeO$\cdot m$Fe$_2$O$_3$ oxides (with w\"ustite, FeO and Fe$_2$O$_3$ as the end-members), and the unclear role of Fe$^{3+}$ in the nature and dynamics of the Earth's lower mantle have attracted much recent attention in geophysics and geochemistry \cite{PhysRevLett.82.4663,PNAS.106.5508,NatCommun.7.10661,Nature.570.102,Jackson_book_1998}.

Under ambient conditions, Fe$_2$O$_3$ crystallizes in a rhombohedral corundum ($R\bar{3}c$ space group) crystal structure with Fe$^{3+}$ cations located in distorted oxygen octahedra \cite{PhysRev.83.333,Greedon_book}. 
It is antiferromagnetic below $\sim$260 K (Morin spin-flop transition temperature) and exhibits weak ferromagnetism (canted antiferromagnetism with a small net ferromagnetic moment of $\sim$0.002 $\mu_B$) between 260 K and the N\'eel temperature of $\sim$956 K \cite{PhysRev.83.333,Greedon_book}.
Photoemission spectroscopy measurements show that Fe$_2$O$_3$ is a Mott-Hubbard insulator with a large energy gap of about 2.5 eV \cite{PhysRevB.34.7318,PhysRevB.39.13478,PhysRevB.66.085115,PhysRevB.79.035108}.
Upon compression above $\sim$50 GPa Fe$_2$O$_3$ undergoes a sharp first-order phase transition to a metallic state (Mott insulator-metal transition) which is accompanied by a remarkable collapse of the lattice volume by about 10\% \cite{PhysScr.43.327,PhysRevLett.82.4663,PhysRevX.8.031059,PhysRevB.65.064112,PhysRevLett.89.205504,JPCM.17.269,JPCS.66.1714,AmMineral.94.205,PNAS.106.5508,PhysRevB.94.014112}. The phase transition has been generally assigned to a high-spin (HS) to low-spin (LS) crossover of Fe$^{3+}$ ions \cite{PhysRevLett.102.146402}, with a complex coexistence (with equal relative abundance) of the non-magnetic and high-spin components in the M\"ossbauer spectra at pressures above 50 GPa \cite{NatCommun.7.10661,PhysRevX.8.031059}. 
It was shown that the transition is accompanied by a structural transformation to the high-pressure crystal structure of Fe$_2$O$_3$. The latter was previously assigned either to an orthorhombic perovskite \cite{PhysScr.43.327} or a Rh$_2$O$_3$-II-type crystal structure ($Pbcn$) \cite{PhysRevLett.82.4663,PhysRevB.65.064112}. 
Only recently, based on single-crystal diffraction, the lattice structure has been resolved to be a distorted double-perovskite GdFeO$_3$-type (DPv) structure (space group $P2_1/n$) \cite{JHPR.33.534,NatCommun.7.10661,PhysRevX.8.031059}.
Furthermore, the fine details of this phase transition seems to depend very much on ``thermal prehistory'' of a sample, showing that the Rh$_2$O$_3$-II-type structure may appear upon heating to about 1800 K with subsequent quenching to low temperatures \cite{NatCommun.7.10661}.
Upon further compression above $\sim$72 GPa, DPv-type Fe$_2$O$_3$ makes a transition to a new high-pressure polymorph whose crystal structure still remains controversial, with two proposed candidates: either a CaIrO$_3$-type post-perovskite (PPv) or orthorhombic $Aba2$ structures \cite{SciRep.5.15091,JHPR.33.534,NatCommun.7.10661,PhysRevX.8.031059}. 

Whereas the electronic properties of the low-pressure corundum ($R\bar{3}c$) phase of Fe$_2$O$_3$ are now well understood from, 
e.g., the LDA+$U$ method \cite{PhysRevB.44.943,PhysRevB.69.165107} (LDA+$U$: density functional theory calculations within the local density approximation plus Hubbard $U$ approach) or the DFT+DMFT calculations \cite{PhysRevLett.62.324,RevModPhys.68.13,RevModPhys.78.865,JPCM.9.7359,PhysRevB.57.6884,EPJST.180.5,PhysRevLett.115.106402} (DMFT: dynamical mean-field theory of correlated electrons), the high-pressure properties of Fe$_2$O$_3$, e.g., its electronic structure, complex coexistence of the HS and LS states observed in the M\"ossbauer spectroscopy, a rich variety of structural polymorph and details of the phase diagram in the megabar pressure range still remain enigmatic \cite{SciRep.5.15091,JHPR.33.534,NatCommun.7.10661,Nature.570.102,PhysScr.43.327,PhysRevLett.82.4663,PhysRevB.65.064112,PhysRevLett.89.205504,JPCM.17.269,JPCS.66.1714,AmMineral.94.205,PNAS.106.5508}. 
Very recently Greenberg \emph{et al.} \cite{PhysRevX.8.031059} detailed the pressure-induced Mott transition in the DPv-type Fe$_2$O$_3$ at about 50 GPa. In our present work, we extend this study focusing on a long-standing challenge of the electronic and magnetic properties of Fe$_2$O$_3$ under ultra-high pressures. We provide a microscopic theory of the high-pressure electronic structure and magnetic state of Fe$_2$O$_3$ up to compression above the core-mantle boundary conditions. Our results reveal that above 75 GPa Fe$_2$O$_3$ adopts a post-perovskite crystal structure, which is characterized by site-selective local moments, with local moments on half of the Fe sites collapsed into the LS state. The Fe $3d$ electrons on the rest of the Fe sites remain 
localized in a high-spin (S=5/2) state up to ultra-high pressures of $\sim$200-250 GPa, well above the core-mantle boundary.
We predict that the site-selective local-moments phase is accompanied by a charge disproportionation of Fe ions, with Fe$^{3\pm \delta}$ and $\delta \sim 0.05$-$0.09$, implying a complex interplay between electronic correlations and the lattice.
 
\section*{Results and Discussion}

\begin{figure}[ht]
\centering
\includegraphics[width=.8\linewidth]{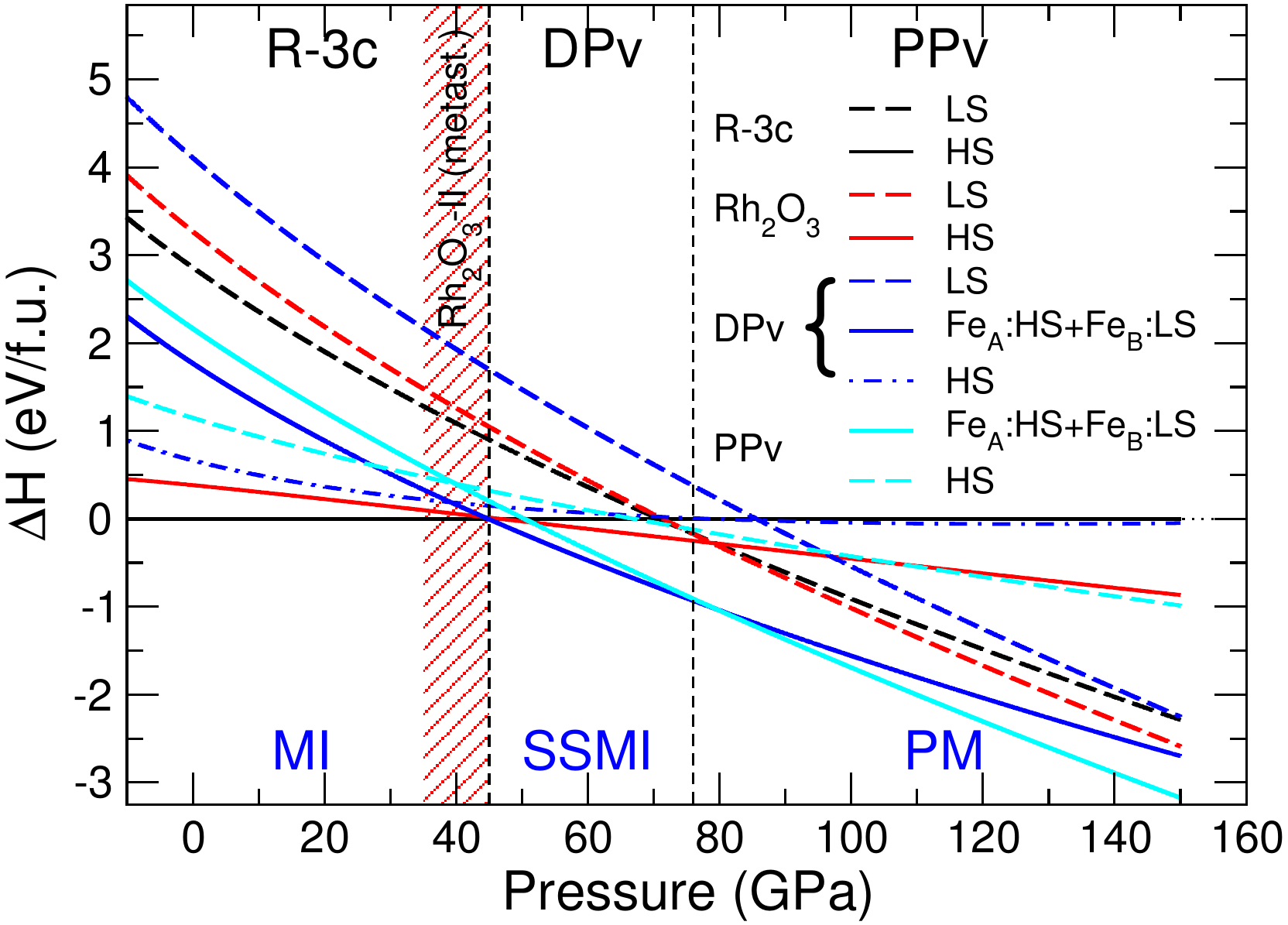}
\caption{Enthalpy difference (relative to the high-spin $R\bar{3}c$ solution) of paramagnetic Fe$_2$O$_3$ calculated by DFT+DMFT as a function of pressure for different phases at $T=1160$~K. MI stands for Mott insulator; SSMI - site-selective MI; PM - paramagnetic metal. Above $\sim$45 GPa, the corundum $R\bar{3}c$ (MI) phase transforms to the double-perovskite DPv (SSMI) phase which above 75 GPa further collapses to the post-perovskite PPv (PM) phase. The corresponding phase transitions are shown by vertical dashed lines. Fe$_\mathrm{A/B}$ are structurally distinct prismatic/octahedral iron sites within the DPv and PPv crystal structures. The metastable Rh$_2$O$_3$-II phase is marked by red shading.}
\label{Fig_1}
\end{figure}

\begin{figure}[ht]
\centering
\includegraphics[width=.8\linewidth]{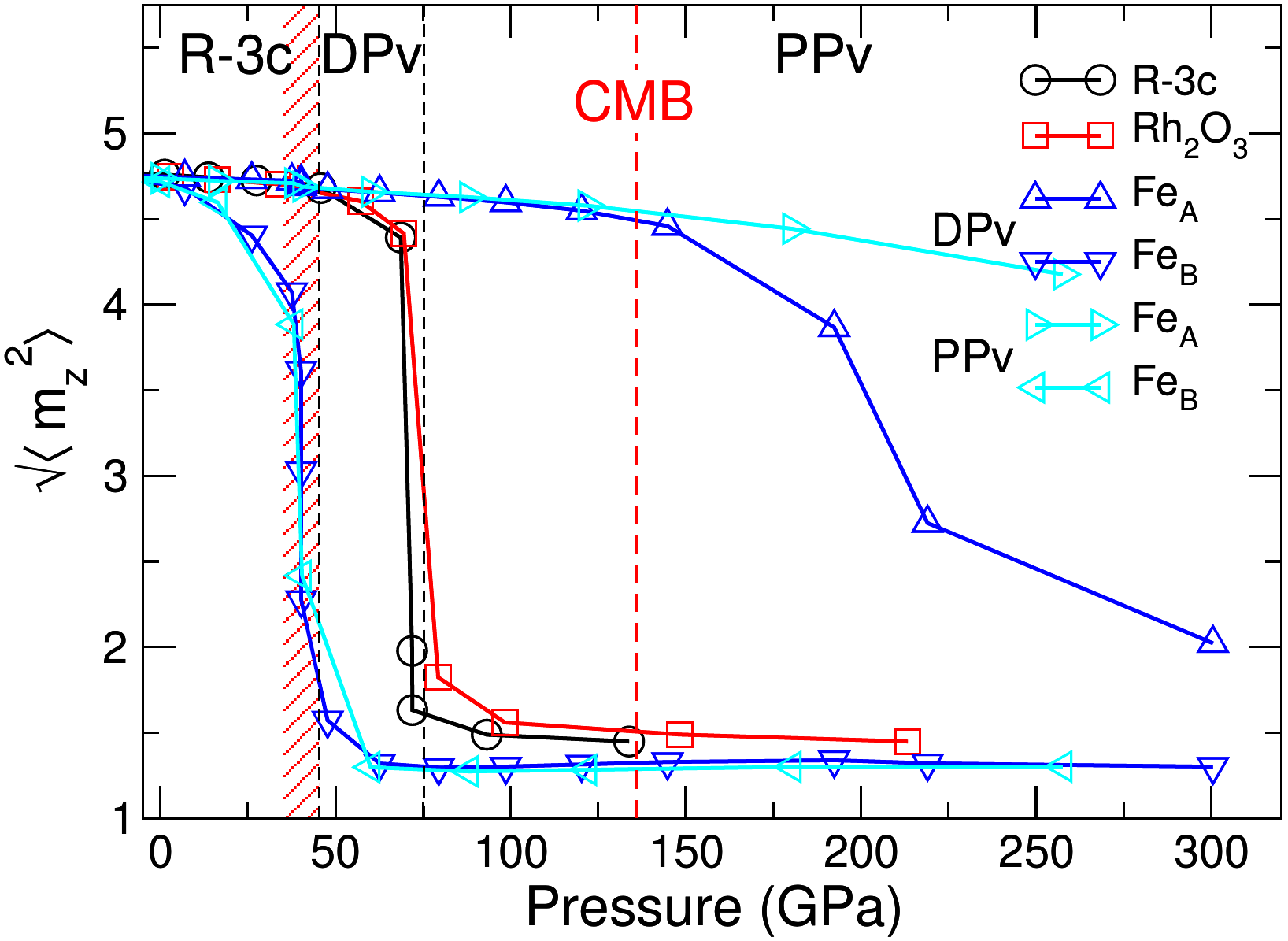}
\caption{Local magnetic moments $\sqrt{ \langle {\hat{m}^2_z} \rangle }$ of paramagnetic Fe$_2$O$_3$ evaluated for different phases within DFT+DMFT as a function of pressure. At high compression, the Fe $3d$ electrons show a complex coexistence of the localized (HS) and delocalized (LS) states, implying a site-selective localized to itinerant moment crossover under pressure. The metastable Rh$_2$O$_3$-II-type phase is marked by red shading. Note that site-selective local moments in PPv Fe$_2$O$_3$ persist up to a high pressure of $\sim$200-250 GPa, well above the core-mantle boundary (CMB).}
\label{Fig_2}
\end{figure}

We employ a state-of-the-art fully self-consistent in charge density DFT+DMFT approach \cite{PhysRevLett.62.324,RevModPhys.68.13,RevModPhys.78.865,JPCM.9.7359,PhysRevB.57.6884,EPJST.180.5,PhysRevB.76.235101,PhysRevB.75.155113,PhysRevB.77.205112,PhysRevB.80.085101,PhysRevB.91.195115} to compute the electronic structure, magnetic state, and crystal structure properties of Fe$_2$O$_3$ under pressure. 
Our results for the calculated enthalpy (with the high-spin $R\bar{3}c$ phase taken as a reference) are summarized in Fig.~\ref{Fig_1}. The pressure-induced evolution of the instantaneous local moments is shown in Fig.~\ref{Fig_2}. Overall, the calculated electronic and lattice properties of Fe$_2$O$_3$ agree well with available experimental data. In agreement with previous studies \cite{PhysRevLett.102.146402,PhysRevX.8.031059}, at ambient pressure, we obtain a Mott insulating solution with a large energy gap of $\sim$2.5 eV 
(see Supplementary Fig.~1). 
The calculated local magnetic moment is $\sim$4.8 $\mu_B$, clearly indicating that at ambient pressure the Fe $3d$ electrons are strongly localized and form a high-spin S=5/2 state (Fe$^{3+}$ ions have a $3d^5$ configuration with three electrons in the $t_{2g}$ and two in the $e_g$ orbitals). 
Our results for the equilibrium lattice constant $a = 5.61$ a.u. and bulk modulus $K_0 \sim 187$ GPa (with $K^\prime \equiv dK/dP$ fixed to 4.1) are in good quantitative agreement with experiment \cite{NatCommun.7.10661,PhysRevX.8.031059}.
Under pressure the energy gap gradually decreases, resulting in a Mott insulator to metal phase transition (MIT), within the $R\bar{3}c$ structure at about 72 GPa, upon compression to below $V \simeq 0.74~V_0$. 
Indeed, our results for the Fe $3d$ spectral function of $R\bar{3}c$ Fe$_2$O$_3$ clearly reveal a sharp increase of the Fe $3d$ density of states at the Fermi level, associated with the Mott MIT (see Supplementary Fig.~2). The phase transition is accompanied by a HS-LS transition, with all the Fe$^{3+}$ ions collapsed to a LS state \cite{PhysRevLett.102.146402}. 
This, however, cannot explain a complex coexistence of the HS and LS Fe$^{3+}$ state near the MIT at pressures above 50 GPa found in the M\"ossbauer experiments (see Refs. \cite{NatCommun.7.10661,PhysRevX.8.031059}, and references therein). In addition, the calculated critical pressure of MIT $\sim$72 GPa is significantly higher, by $\sim$40 \%, than that found in the experiments \cite{PhysRevLett.82.4663,PhysRevX.8.031059}. This manifests crucial importance of the interplay of the electronic state and lattice near the Mott transition in Fe$_2$O$_3$. In fact, it is known experimentally that the MIT in Fe$_2$O$_3$ is accompanied by  a structural transformation to a double perovskite GdFeO$_3$-type 
or Rh$_2$O$_3$-II-type ($Pbcn$) structure above $\sim$50 GPa. 

We compute the crystal structure phase stability and the electronic structure of the DPv and Rh$_2$O$_3$-II-type Fe$_2$O$_3$. To explore the phase stability of Fe$_2$O$_3$ near the Mott transition, we use the atomic positions and the crystal structure parameters taken from the experiments at about 50 GPa \cite{NatCommun.7.10661,PhysRevX.8.031059}. We use the DFT+DMFT approach with two (inequivalent) impurity sites in the unit cell in order to treat electron correlations in the Fe $3d$ bands of the structurally distinct Fe $A$ (which have a prismatic oxygen coordination) and Fe $B$ (octahedral) sites of the DPv, as well as of the PPv phase.
We note that the DPv phase of Fe$_2$O$_3$ (monoclinic space group $P2_1/n$) adopts a general formula A$_2B'B''O_6$, i.e., it contains two crystallographically nonequivalent Fe $B$ sites (with octahedral oxygen coordination). Our DFT+DMFT calculations with explicit treatment of the crystallographically different Fe $A$, $B'$ and $B''$ sites (three-site impurity problem within DMFT) show that the Fe $B'$ and $B''$ ions have very similar electronic state: similar spectral functions (in the local coordinate system), occupations, and
local magnetic moments. Because of this we average the Green's functions corresponding to the Fe $B'$ and $B''$ sites in our total-energy calculations and solve two-site impurity problem within DMFT (one for the Fe $A$ and one for the averaged Fe $B$ site).

In agreement with experiment, the $R\bar{3}c$ phase is found to be energetically favorable at ambient pressure, with a total-energy difference of about 670 meV/f.u. and 350 meV/f.u. between corundum and the DPv, and corundum and the Rh$_2$O$_3$-II-type structures, respectively. 
Our results show that above $\sim$45 GPa (upon compression below $\sim 0.84~V_0$), Fe$_2$O$_3$ undergoes a phase transition from corundum to the double-perovskite DPv crystal structure (see Fig. \ref{Fig_1}). The Rh$_2$O$_3$-II lattice is energetically unfavorable, i.e., thermodynamically unstable, with a total energy difference with respect to the DPv phase of $\Delta E \sim 42$ meV/f.u. for $V=0.84~V_0$. Our result therefore suggests that at high temperatures the Rh$_2$O$_3$-II-type crystal structure can become metastable prior to the transition to the high-pressure DPv Fe$_2$O$_3$, in agreement with recent experiments \cite{JHPR.33.534,NatCommun.7.10661}. Interestingly, both the $R\bar{3}c$ and Rh$_2$O$_3$-II phases exhibit a Mott MIT, accompanied by the HS-LS transition near 72 GPa.
Most importantly, the $R\bar{3}c$-to-DPv phase transition is found to take place at $\sim$45 GPa, in quantitative agreement with experiment \cite{NatCommun.7.10661,PhysRevX.8.031059}. This clearly demonstrates the crucial importance of the interplay of electronic correlations and the lattice to explain the properties of Fe$_2$O$_3$. The phase transition is accompanied by a collapse of the lattice volume by $\sim$11.6 \%, whereas 
the calculated bulk modulus of the DPv phase ($\sim$259 GPa) is found to be substantially larger than that obtained for the HS state of the $R\bar{3}c$ phase (187 GPa). 

\begin{figure}[ht]
\centering
\includegraphics[width=.8\linewidth]{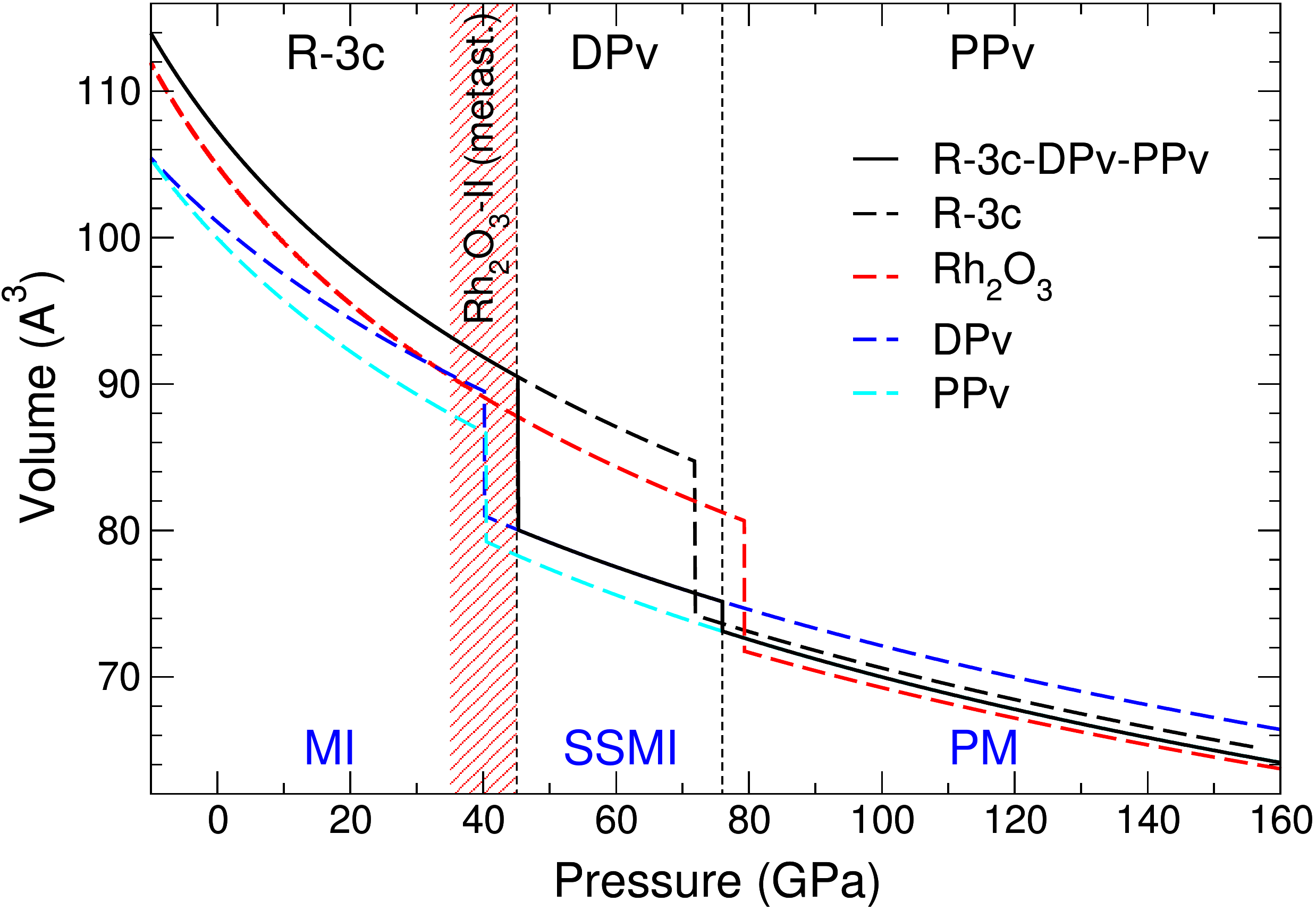}
\caption{Equation of state of Fe$_2$O$_3$ calculated by DFT+DMFT for different phases at $T=1160$~K. Fe$_2$O$_3$ undergoes the corundum ($R\bar{3}c$) to double-perovskite (DPv) phase transition at about 45 GPa, and then above 75 GPa, it makes a DPv-PPv  transformation (PPv - postperovskite). MI stands for Mott insulator, SSMI for site-selective MI, and PM for paramagnetic metal. The metastable Rh$_2$O$_3$-II-type phase is marked by red shading.}
\label{Fig_3}
\end{figure}

Furthermore, we examine the high-pressure electronic structure and phase stability of paramagnetic Fe$_2$O$_3$ in the orthorhombic $Aba2$ and the CaIrO$_3$-type post-perovskite (PPv) crystal structures, i.e., for the two structural candidates for the high-pressure metallic phase proposed from experiment \cite{JHPR.33.534,NatCommun.7.10661,PhysRevX.8.031059}. Our total-energy calculations within a single-site DFT+DMFT method for the $Aba2$ phase reveal its remarkable thermodynamic instability (as high as $\sim$3.71 eV/f.u. above the PPv phase).
We note that while this value is large, $\sim$740 meV/atom, it is not unrealistic. Here, we refer to two new metastable phases of SiO$_2$, coesite-IV and coesite-V, which have been synthesized experimentally \cite{NatCommun.9.4789}. The phases are energetically highly unfavorable -- at 38 GPa, where coesite-V and coesite-IV are nearly degenerate in enthalpy in theoretical calculations, the calculated enthalpy difference between them and the ground state was found to be $\sim$390 meV/atom. The values calculated for carbon polymorphs, for example, such as diamond and C$_{60}$ are even higher.
DFT+DMFT calculations with different computational parameters ($U$ and $J$) show that the obtained result is robust, suggesting that $Aba2$ Fe$_2$O$_3$ is metastable at high pressures. Our result agrees well with recent x-ray diffraction studies which reveal that the $Aba2$ phase is in fact metastable with a stability range limited to low-temperatures \cite{NatCommun.7.10661}. Moreover, the latter is experimentally found to transform into the PPv structure upon annealing to high temperatures \cite{NatCommun.7.10661}. 
Interestingly, the crystal structure analysis of the $Aba2$ structure of Fe$_2$O$_3$ reveals the existence of two sufficiently different Fe-Fe bond distances: the Fe-Fe pairs with a short inter-atomic distance of $\sim$2.43 \AA\ and the rest with that of 2.73 \AA. While the $Aba2$ phase is non-magnetic (according to the low-temperature M\"ossbauer spectroscopy \cite{PhysRevX.8.031059}), this may suggest dimerization of the Fe-Fe ions in $Aba2$ Fe$_2$O$_3$. As a consequence, it implies a possible importance of non-local correlation effects to explain the appearance of the metastable $Aba2$ phase at low temperatures. 
Based on our theoretical results, we predict a structural phase transition from the DPv to PPv phase above $\sim$75 GPa, in quantitative agreement with available experiments \cite{NatCommun.7.10661,PhysRevX.8.031059}. The phase transition is accompanied by about 2.6 \% collapse of the lattice volume and is associated with formation of a metallic state (both Fe $A$ and $B$ sublattices are metallic). 
In Fig. \ref{Fig_3} we summarize our results for the equation of state of the Fe$_2$O$_3$ polymorphs considered in the present study, revealing two distinct anomalies at about 45 GPa and $\sim$75 GPa.

Our results for the local magnetic moments and spectral properties confirm that a structural transition from the low-pressure corundum $R\bar{3}c$ to the DPv structure is accompanied by a Mott insulator to site-selective MI phase transition, in agreement with recent experiments \cite{PhysRevX.8.031059}. 
In fact, the phase transition above 45 GPa is characterized by a collapse of local moments 
and emergence of a metallic state only on half of the Fe sites (see Fig.~\ref{Fig_2} and Supplementary Figs.~3 and 4). The spectral function of the Fe $B$ ions exhibits a large increase of its weight at the Fermi level for $V<0.85~V_0$, whereas the Fe $A$ spectral weight at $\varepsilon_F$ remains unchanged down to $\sim$0.6~V/V$_0$. That is, the prismatic Fe $A$ sites remain insulating with the local magnetic moment of $\sim$4.6 $\mu_B$ (high-spin, S=5/2), which is substantially larger than that for the Fe $B$ sites 0.89 $\mu_B$ (low-spin, S=1/2) at about 70 GPa.
The HS-LS transition of the Fe $A$ electrons in DPv Fe$_2$O$_3$ is found to occur at a  substantially higher compression of $\sim 0.6$ $V_0$, above 192 GPa, and is accompanied by $\sim$5 \% collapse of the lattice volume. The latter transition is associated with a formation of a fully metallic state of the DPv Fe$_2$O$_3$ phase, where both the Fe $A$ and $B$ sublattices are metallic.

\begin{figure}[ht]
\centering
\includegraphics[width=.95\linewidth]{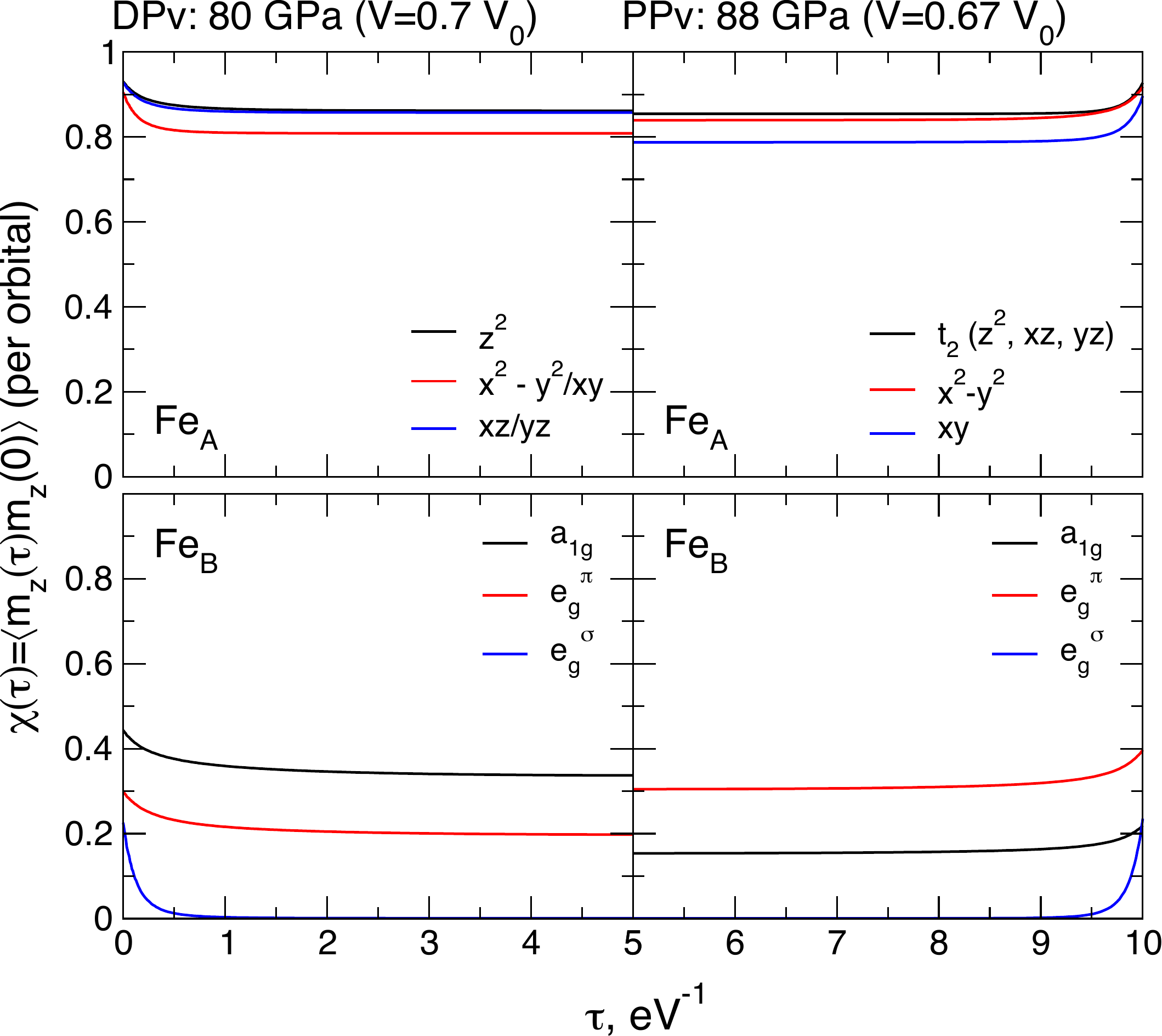}
\caption{Local spin susceptibility $\chi=\langle \hat{m}_z(\tau)\hat{m}_z(0) \rangle$ calculated by DFT+DMFT for the paramagnetic DPv (left panel) and PPv (right panel) phases of Fe$_2$O$_3$, for $\sim$80 GPa (for the DPv phase) and 88 GPa (PPv), respectively. The Fe $A$ electrons (associated with the prismatic Fe $A$ sites) are localized to form fluctuating moments, while the octahedral Fe $B$ sites are delocalized (itinerant-like). $\tau$ denotes the imaginary time, $\tau \in [0,\beta]$, where $\beta$ is the inverse temperature, $\beta=1/k_BT$.}
\label{Fig_4}
\end{figure}

Our calculations show that Fe$_2$O$_3$ in the PPv phase exhibits a site-selective local moments behavior similar to that found in the SSMI DPv phase with the local magnetic moment of $\sim$4.58$\mu_B$ for the Fe $A$ sites (with prismatic oxygen coordination), which is substantially larger than that for the octahedral Fe $B$ sites $\sim$0.97$\mu_B$,  around 88 GPa (see Fig. \ref{Fig_2}).
Our calculations predict that site-selective local moments within the PPv phase persist up to a very high compression of $\sim$200-250 GPa. That is, even at pressures near the core mantle boundary ($\sim$140 GPa) Fe$_2$O$_3$ remains highly (para-) magnetic. The latter is particularly remarkable because both the prismatic Fe $A$ and the octahedral Fe $B$ sublattices are metallic.
Moreover, analysis of the local (dynamical) spin-spin correlation function $\chi(\tau)=\langle \hat{m}_z(\tau) \hat{m}_z(0) \rangle$ of the DPv and PPv Fe$_2$O$_3$ shown in Fig.~\ref{Fig_4} reveals that in both phases the Fe $A$ electrons are localized to form fluctuating moments [$\chi(\tau)$ is seen to be almost constant, independent of $\tau$, and close to the unit], while the Fe $B$ electrons are delocalized (itinerant-like). 
Our findings, therefore, suggest that the MIT in the DPv and PPv Fe$_2$O$_3$ under pressure is associated with a site-selective delocalization of the Fe $3d$ electrons.
We note that similar behavior associated with the crossover from a localized to itinerant state has recently been shown to appear at the pressure-induced Mott transition in monoxides MnO, FeO, CoO, and NiO \cite{PhysRevB.92.085142,PhysRevB.94.155135,PhysRevB.96.075136}.

Our results  provide a microscopic understanding of the complex coexistence of the HS and LS states observed in the high-pressure M\"ossbauer spectroscopy. In fact, this state can be rationalized as a site-selective Mott state of Fe$_2$O$_3$ in which local moments on half of the Fe sites collapse into the LS (metallic) state. 
Interestingly, our results suggest that the HS-LS state transition is accompanied by a remarkable enhancement of the crystal-field splitting (both in the DPv and PPv phases), caused by correlation effects (see Supplementary Fig.~5 and discussion there). We find that the crystal field energy splitting $\Delta_{cf}$ for the Fe $3d$ states (determined from the first moments of the interacting lattice Green's function) is larger for the small-moment Fe $B$ sites with respect to the large-moment Fe $A$ ions. Upon compression above the critical value, $\Delta_{cf}$ enhances by a factor of $\sim$2, resulting in the site-selective HS-LS transformation.
Moreover, our results show sufficiently different hybridization strength for the low-spin Fe $B$ sites with respect to that in the high-spin Fe $A$ site, which seems to mediate the critical value of the crystal-field splitting determined from the Hund's coupling $J$ energy value (see Supplementary Fig.~6).

\begin{figure}[ht]
\centering
\includegraphics[width=.8\linewidth]{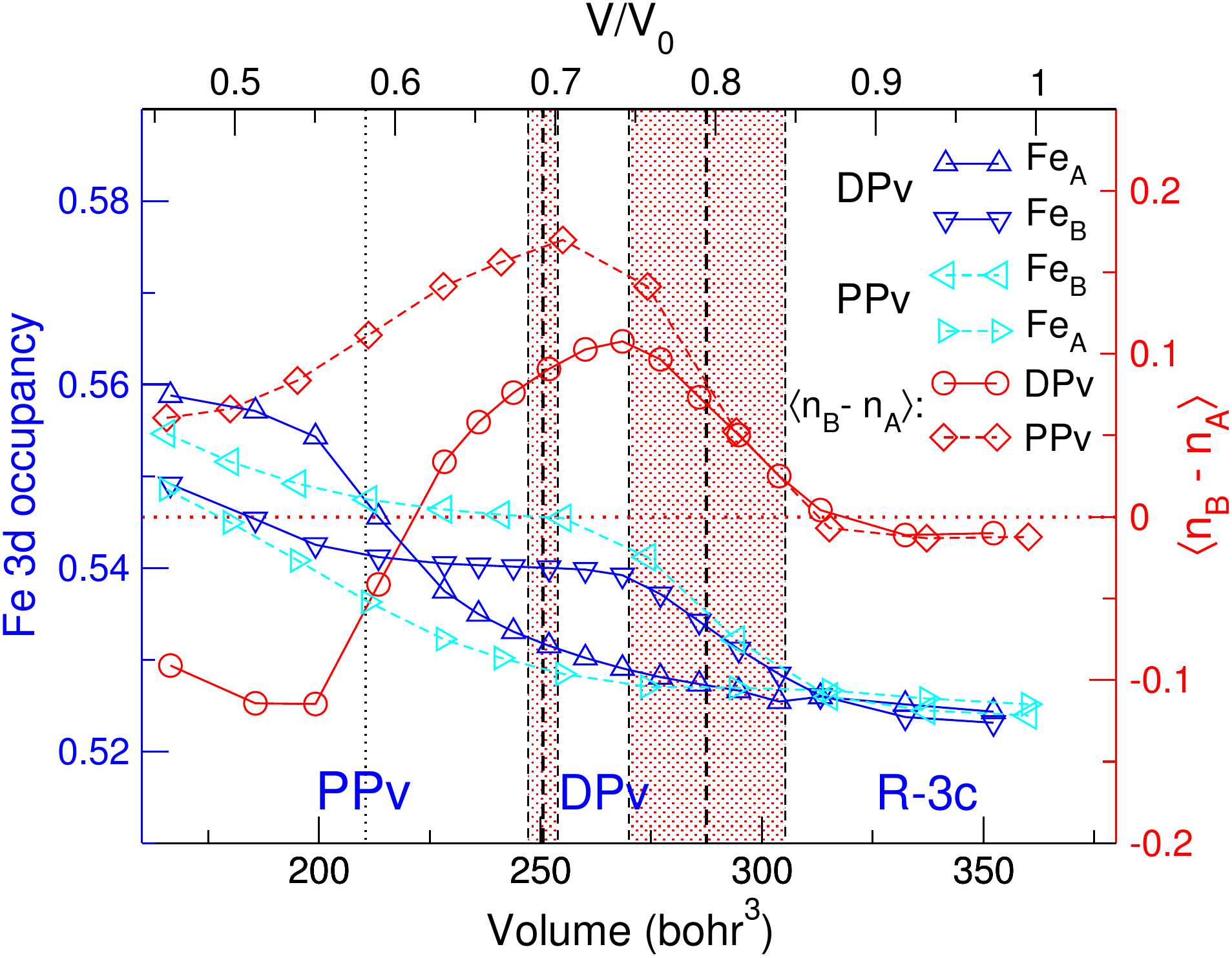}
\caption{Evolution of the Wannier Fe $3d$ total occupations and Fe$^{3 \pm \delta}$ charge disproportionation with $\delta = (\langle \hat{n}_B \rangle - \langle \hat{n}_A\rangle)/2$ obtained by DFT+DMFT for the DPv and PPv phases as a function of volume. The $R\bar{3}c$-DPv and DPv-PPv phase transitions are shown by vertical dashed lines. Our results for the lattice volume collapse are marked by red shading. The SSMI-PM phase transition within the DPv crystal structure (at $\sim$0.58~V$_0$, corresponding to 192 GPa) is marked by a dotted line.}
\label{Fig_5}
\end{figure}

Interestingly, our results for both the DPv and PPv phases predict a charge disproportionation between the Fe $A$ and $B$ sites in Fe$_2$O$_3$. In Fig. \ref{Fig_5} we show the total Fe $A$ and $B$ $3d$ occupations and Fe$^{3 \pm \delta}$ $3d$ charge difference with $\delta \equiv \langle \hat{n}_{\mathrm{B}}-\hat{n}_{\mathrm{A}} \rangle/2$ as a function of volume. In particular, while it is seen to be negligible in the \emph{conventional} Mott phase (MI) of the DPv and PPv Fe$_2$O$_3$ 
(i.e., it is absent even despite the fact that the $A$ and $B$ sites have very different oxygen environment, $\delta \sim 0$),
the charge difference increases upon transition into the SSMI phase, and is about $\delta \sim0.05$ and 0.09 for the DPv and PPv phases, respectively. Note that the charge disproportionation is absent in the $R\bar{3}c$ and Rh$_2$O$_3$-II phases (by symmetry all crystallographic positions of Fe are equivalent). Moreover, it is interesting to point out that the corresponding $3d$ charge difference ($2\delta=0.1$-0.18) is in agreement with a charge disproportionation value of $\sim$0.2 found in the low-temperature charge-ordered phases of the mixed-valent Fe-based oxides, such as Fe$_3$O$_4$ and Fe$_2$BO$_4$ \cite{PhysRevB.66.214422,Nature.481.173, NatCommun.10.2857,1367-2630-7-1-053,PhysRevB.74.165117,PhysRevB.74.195115,doi:10.1063/1.3584855,PhysRevB.72.014407}. 

We also notice that the sign of $\delta$ is opposite to what is expected: the small-volume octahedral Fe $B$ cations, appear to hold more electrons than the prismatic Fe $A$, located in the largest oxygen cage. This  
behavior is a consequence of emptying of the antibonding $e_g^\sigma$ states of the octahedral Fe $B$ sites at the MI to SSMI phase transition, which leads to a different strength of covalent $p$-$d$ bonding for the Fe $A$ and $B$ sites \cite{PhysRevLett.106.256401}.
Moreover, upon further compression of the lattice below $\sim$0.6 V$_0$, above $\sim$190 GPa, the charge disproportion is found to change sign upon transition to a (conventional) metallic state in the DPv phase. In the PPv Fe$_2$O$_3$ it tends to decrease below $\delta \sim 0.05$, implying a complex interaction between electronic correlations, local magnetism, and the lattice on microscopic level. We note that a similar behavior has been recently suggested to occur in nanocrystalline Fe$_2$O$_3$ under pressure \cite{PhysRevB.86.205131}. In addition, we point out a similarity of the electronic and magnetic behavior of Fe$_2$O$_3$ to that observed in the rare-earth nickelates ($R$NiO$_3$ with $R$=Sm, Eu, Y, or Lu) \cite{PhysRevLett.109.156402,PhysRevB.91.075128,PhysRevB.92.155145,PhysRevB.96.205139,PhysRevB.96.045120}. 
In fact, the latter exhibit a site-selective Mott transition, characterized by a two-sublattice symmetry breaking, with formation of site-selective local moments with localized Ni $3d$ magnetic moments and Ni-$d$--O-$p$ singlet nonmagnetic states. The localized Ni $3d$ moments are formed on the nickel sublattice having an increased mean Ni-O bond, while the nonmagnetic states have a decreased Ni-O bond.
Note that the nickelates remain insulating while exhibiting formation of site-selective local moments, in qualitative difference from a site-selective Mott-insulating state of Fe$_2$O$_3$. Moreover, a charge disproportionation between the local moment and Ni-$d$--O-$p$ singlet sites is also small, $2\delta =0.1$-0.16 electrons, in accordance with our result of $\sim$0.13-0.18 for Fe$_2$O$_3$ \cite{PhysRevLett.109.156402,PhysRevB.91.075128}. Because of this we suggest that charge disproportionation is the intrinsic property of the site-selective Mott phase of Fe$_2$O$_3$, which appears due to the complex interplay between electronic, magnetic, and lattice degrees of freedom. In fact, the LS state of Fe$^{3+}$ ions experience more charge transfer from the ligands, resulting in a sufficient charge disproportionation between the HS and the LS states in the SSMI phase.

Overall, our calculations show that under pressure Fe$_2$O$_3$ undergoes a series of complicated electronic, magnetic, and structural transformations. It starts with a phase transition from the corundum to DPv phase at about 45 GPa, and then above 75 GPa, Fe$_2$O$_3$ makes a DPv-PPv transformation. 
All of these phase transformations are of first order, resulting in a substantial change of the lattice volume by $\sim$2.6$-$11.6 \% (see Fig.~\ref{Fig_3}).
Our calculations reveal a complex interplay of electronic correlations and the crystal structure near the Mott transition, resulting in the formation of a site-selective Mott insulating phase and site-selective collapse of the local moments. 
Below a megabar, our results show a remarkable competition of different crystallographic phases of
Fe$_2$O$_3$, suggesting the importance of  metastable phases for understanding its crystallochemistry.
Most importantly, our results predict that site-selective local moments in the metallic PPv Fe$_2$O$_3$ persist up to ultra-high compression of $\sim$200-250 GPa, i.e., well above the core-mantle boundary pressure.
This, in fact, can be important for the understanding of the velocity and density anomalies in the Earth's lower mantle, providing a possible explanation for the origin of large low-shear-velocity provinces.

In conclusion, we show that the interplay between electronic correlations and the lattice in the vicinity of a Mott transition results in the formation of complex electronic and magnetic states in Fe$_2$O$_3$. This gives rise to a remarkable structural complexity of Fe$_2$O$_3$, implying a possible importance of metastable structures for understanding its high-pressure and high-temperature behavior. 
Our results explain microscopically the coexistence of the HS and LS states observed in M\"ossbauer spectroscopy at high pressures, suggesting the existence of a novel class of Mott systems -- with site-selective local moments -- which may have important impact for understanding the properties and evolution of the Earth's lower mantle and outer core.
We expect that the electronic and structural complexity of Fe$_2$O$_3$ under pressure revealed in this study, e.g., its complex allotropy and the presence of metastable phases may affect present geophysical and geochemical models. 

\section*{Methods}

We have employed the DFT+DMFT approach to explore the electronic structure, local magnetic state of Fe$^{3+}$ ions, and crystal structure stability of paramagnetic Fe$_2$O$_3$ under pressure 
using the DFT+DMFT method \cite{PhysRevB.76.235101,PhysRevB.75.155113,PhysRevB.77.205112,PhysRevB.80.085101,PhysRevB.91.195115} implemented with plane-wave pseudopotentials \cite{JPhysCondMat.21.395502,PhysRevLett.101.096405,PhysRevB.81.075109}.
We start by constructing the effective low-energy O $2p$ - Fe $3d$ Hamiltonian [$\hat{H}^{\mathrm{DFT}}_{\sigma,\alpha\beta}(\bf{k})$] using the projection onto Wannier functions to obtain the $p$-$d$ Hubbard Hamiltonian (in the density-density approximation)
\begin{eqnarray}
\label{eq:hamilt}
\hat{H} = \sum_{\bf{k},\sigma} \hat{H}^{\mathrm{DFT}}_{\sigma,\alpha\beta}({\bf{k}}) + \frac{1}{2} \sum_{i,\sigma\sigma',\alpha\beta} U_{\alpha\beta}^{\sigma\sigma'} \hat{n}_{i,\alpha\sigma} \hat{n}_{i,\beta\sigma'} - \hat{H}_{\mathrm{DC}},
\end{eqnarray}
where $\hat{n}_{i,\alpha\sigma}$ is the occupation number operator for the $i$-th Fe site with spin $\sigma$ and (diagonal) orbital indices $\alpha$.
For this purpose, for the partially filled Fe $3d$ and O $2p$ orbitals we construct a basis set of atomic-centered symmetry-constrained Wannier functions \cite{RevModPhys.84.1419,PhysRevB.71.125119,JPhysCondMat.20.135227}. 
The Wannier functions are constructed using the scheme of Refs.~\cite{PhysRevB.71.125119,JPhysCondMat.20.135227}: the O $2p$ orbitals were constructed using Wannier functions defined over the full energy range spanned by the $p$-$d$ band complex; the localized Fe $3d$ orbitals are constructed using the Fe $3d$ band set. 
All the calculations are performed in the local basis set determined by diagonalization of the corresponding Fe $3d$ occupation matrices.
In order to solve the realistic many-body problem, we employ the continuous-time hybridization-expansion quantum Monte-Carlo algorithm \cite{RevModPhys.83.349}.
The Coulomb interaction has been treated in the density-density approximation. The elements of the $U$ matrix are parametrized by the average Coulomb interaction $U$ and Hund's exchange $J$ for the Fe $3d$ shell. For all the structural phases considered here we have used the same $U=6$ eV and $J=0.86$ eV values as was estimated previously \cite{PhysRevLett.102.146402,PhysRevX.8.031059}. 
The spin-orbit coupling was neglected in these calculations. Moreover, the $U$ and $J$ values are assumed to remain constant upon variation of the lattice volume.
We employ the fully localized double-counting correction, evaluated from the self-consistently determined local occupations, to account for the electronic interactions already described by DFT,
$\hat{H}_{DC}=U ( N - \frac{1}{2} ) - J ( N_{\sigma} - \frac{1}{2} )$,
where $N_\sigma$ is the total Fe $3d$ occupation with spin $\sigma$ and $N=N_\uparrow+N_\downarrow$.
Here, we employ a fully self-consistent in charge density DFT+DMFT scheme in order to take into account the effect of charge redistribution caused by electronic correlations and electron-lattice coupling.
The spectral functions were computed using the maximum entropy method.

We explore the structural stability of Fe$_2$O$_3$ up to a pressure of 200 GPa, and, in particular, the influence of electronic correlations on the structural transition(s) and the electronic state of iron ions. 
To this end, we adopt the crystal structure data for the corundum ($R\bar{3}c$), double-perovskite ($P2_1/n$), Rh$_2$O$_3$-II-type ($Pbcn$), post-perovskite ($Cmcm$), and orthorhombic ($Aba2$) structures taken from experiment \cite{NatCommun.7.10661,PhysRevX.8.031059}, and calculated enthalpy within DFT+DMFT. To compute pressure we fit the calculated total energies using the third-order Birch-Murnaghan equation of states separately for the low- and high-volume regions.

\section*{Data availability}

The data that support the findings of this study are available from the corresponding authors upon  request.

\section*{Code availability}

The DFT+DMFT code employed in this study is available from the corresponding authors upon  request.

\section*{Acknowledgments}

We thank E. Greenberg, L. Dubrovinsky, and R. Jeanloz for valuable discussions. Theoretical analysis of structural properties was supported by the Russian Science Foundation (Project No. 18-12-00492). Support provided by the Swedish Research Council project No. 2015-04391, the Swedish Government Strategic Research Areas in Materials Science on Functional Materials at Link\"oping University (Faculty Grant SFO-Mat-LiU No. 2009-00971) and the Swedish e-Science Research Centre (SeRC) is gratefully acknowledged. This research was supported in part by Israeli Science Foundation Grant \#1189/14 and \#1552/18.

\section*{Competing Interests}

The authors declare that they have no competing interests.

\section*{Author contributions}

I.L. performed the theoretical analysis; all the authors (I.L., G.Kh.R., and I.A.A.) contributed to the interpretation of the data and to the writing of the manuscript.

\section*{Corresponding authors}
	
Correspondence to Ivan Leonov or Igor A. Abrikosov.

\newpage

\subsection*{Figures}

\pagebreak

\pagebreak

\pagebreak

\pagebreak

\newpage
\section*{Supplementary Information}

In our DFT+DMFT calculations we used the average Coulomb interaction $U = 6$ eV and Hund's exchange $J = 0.86$ eV for the Fe $3d$ shell as was estimated previously \cite{PhysRevLett.102.146402,PhysRevX.8.031059}. In order to check how the DFT+DMFT results depend on the choice of the Hubbard $U$ parameter we compute the electronic structure and magnetic state of $R\bar{3}c$ Fe$_2$O$_3$ for the Hubbard $U$ values varied from $U=5$ eV to $U=8$ eV. The Hund's coupling was fixed ($J=0.86$ eV). Our results are shown in Supplementary Fig.~\ref{Fig_S1}. For this broad range of the Hubbard $U$ values our results exhibit no sufficient modification of the electronic structure and magnetic state of Fe$_2$O$_3$. We obtain that the local magnetic moments slightly increase from 4.71 $\mu_B$ ($U=5$ eV) to 4.78 $\mu_B$ ($U=8$ eV). The energy gap however depends sensitivity on the choice of the $U$ value, increasing by $\sim$0.9 eV from 2.03 eV for $U=5$ eV to 2.95 eV for $U=8$ eV. 
At ambient pressure, for $R\bar{3}c$ Fe$_2$O$_3$ (with $U=6$~eV and $J=0.86$~eV) we obtain a Mott insulating solution with a relatively large $d$-$d$ energy gap of about 2.5 eV. Under pressure, the energy gap is found to gradually decrease, and it is about 1.9 eV at a pressure of $\sim$45 GPa.

\begin{figure}[ht]
\centering
\includegraphics[width=1.0\linewidth]{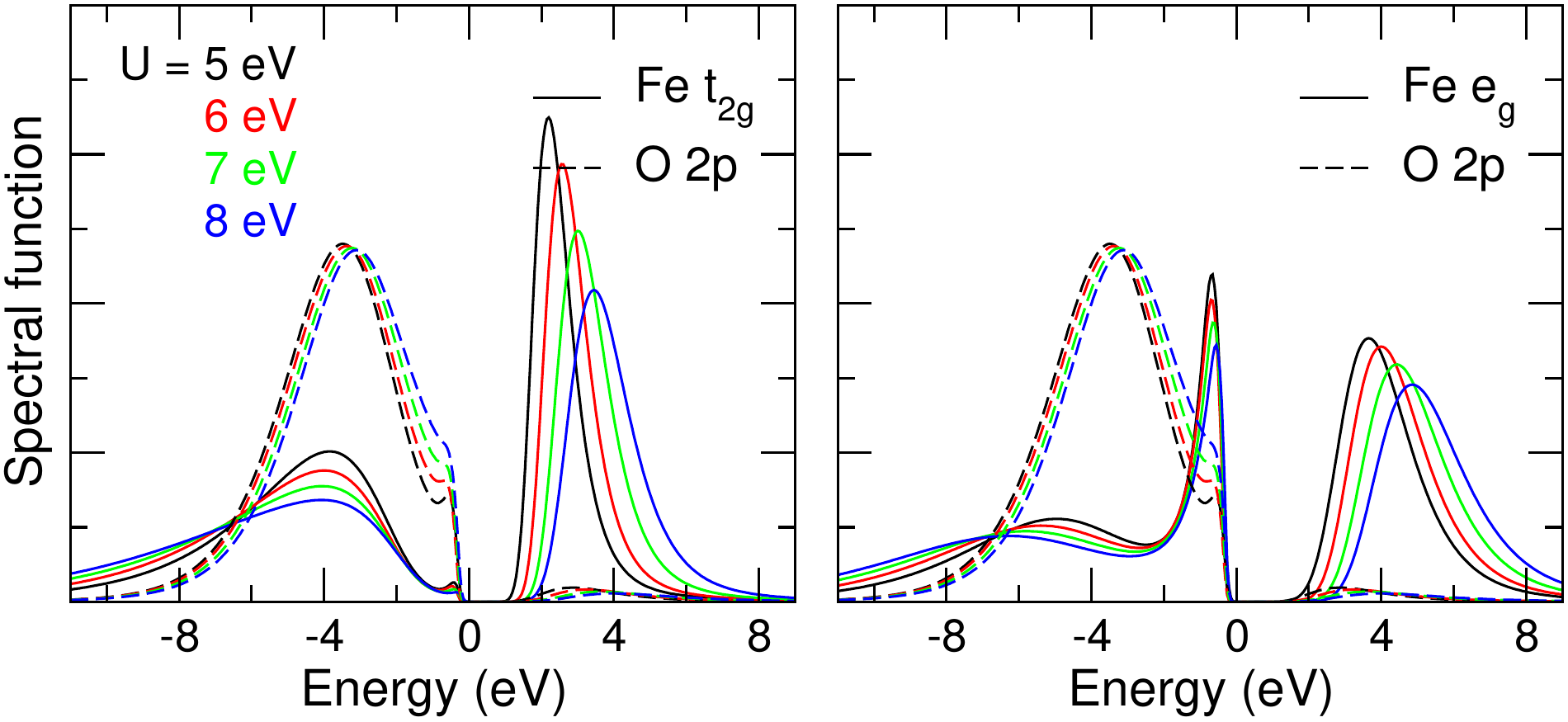}
\caption{Evolution of the Wannier Fe $3d$ and O $2p$ spectral functions of $R\bar{3}c$ Fe$_2$O$_3$ obtained by DFT+DMFT for the Hubbard $U$ values varied from $U=5$ eV to $U=8$ eV. The Hund's coupling is fixed to $J=0.86$ eV. 
}
\label{Fig_S1}
\end{figure}

We compute a pressure-induced evolution of the spectral weight at the Fermi level $N(\varepsilon_F) = -\frac{\beta}{\pi} G( \tau =\frac{\beta}{2})$ for the Fe $3d$ orbitals in paramagnetic Fe$_2$O$_3$ using the DFT+DMFT method.  Here, $G(\tau)$ is the interacting Green's function on the imaginary-time $\tau$ domain, evaluated within DMFT. $\beta$ is the inverse temperature, $\beta=1/k_BT$. Our results for the $R\bar{3}c$, DPv, and PPv crystal structures of Fe$_2$O$_3$ are summarized in Supplementary Fig.~\ref{Fig_S2}. Our results clearly reveal a sharp increase of the Fe $3d$ density of states at the Fermi level, associated with a change of their electronic state. In particular, in Fe$_2$O$_3$ with the $R\bar{3}c$ crystal lattice  $N(\varepsilon_F)$ greatly enhances upon compression below $V \sim 0.74~V_0$, above 72 GPa, resulting in metallization (i.e., in the Mott MIT). The phase transition is accompanied by a HS-LS transition, with all the Fe$^{3+}$ ions collapsed to a
LS state. In the DPv phase metallization of the prismatic Fe $A$ and octahedral Fe $B$ (consists of the $B'$ and $B''$ sites) sublattices occurs upon two sufficiently different compressions. First, the Fe $B$ sublattice becomes metallic under pressure below $\sim$0.8 $V_0$ (above 40 GPa) and then upon substantially higher compression of $\sim$0.6 $V_0$ (above 192 GPa), the Fe A $3d$ ions become metallic. In the PPv phase we observe the formation of a site-selective HS/LS state under pressure above $\sim$40 GPa, below $\sim$0.78~V$_0$.

\begin{figure}[ht]
\centering
\includegraphics[width=.7\linewidth]{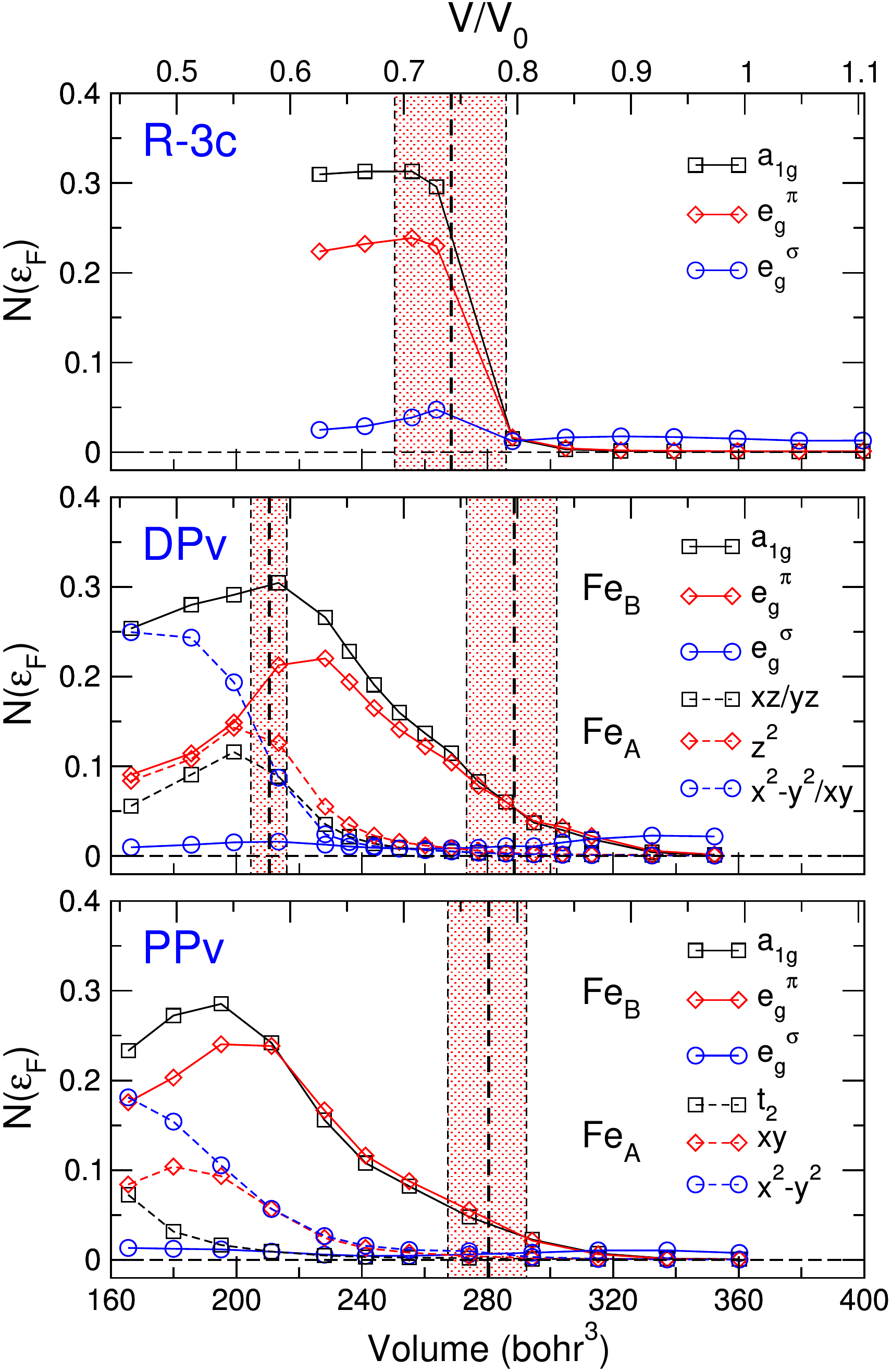}
\caption{Evolution of the Wannier Fe $3d$ and O $2p$ spectral weight $N(\varepsilon_F)$ at the Fermi level obtained by DFT+DMFT for the $R\bar{3}c$, DPv and PPv phases as a function of volume. Our results for the lattice volume collapse associated with a Mott to site-selective Mott phase (site-selective HS-LS state) are marked by red shading. Note two consecutive phase transformations in the DPv phase:
the Mott insulator (MI) to site-selective MI (SSMI) phase transition below $\sim$0.8 V$_0$ (40 GPa) and the SSMI to paramagnetic metal phase transformation below $\sim$0.6 V$_0$. }
\label{Fig_S2}
\end{figure}

In Supplementary Fig.~\ref{Fig_S3}, we present our results for the spectral function of the DPv phase of Fe$_2$O$_3$ calculated using DFT+DMFT for $V/V_0=0.71$ ($\sim$72 GPa). Our results show the existence of a site-selective Mott insulating phase, in which the $3d$ electrons on only half of the Fe sites (octahedral $B$ sites) are metallic (i.e., delocalized), while the prismatic Fe $A$ sites are still insulating (Fe $3d$ electrons are localized). The spectral function of PPv Fe$_2$O$_3$ at a high pressure of $\sim$122 GPa ($\sim$0.63 $V_0$) is shown in Supplementary Fig.~\ref{Fig_S4}. Under high compression, DFT+DMFT gives a metallic state, with both crystallographic Fe $A$ and $B$ sites being a metal.

\begin{figure}[ht]
\centering
\includegraphics[width=1.0\linewidth]{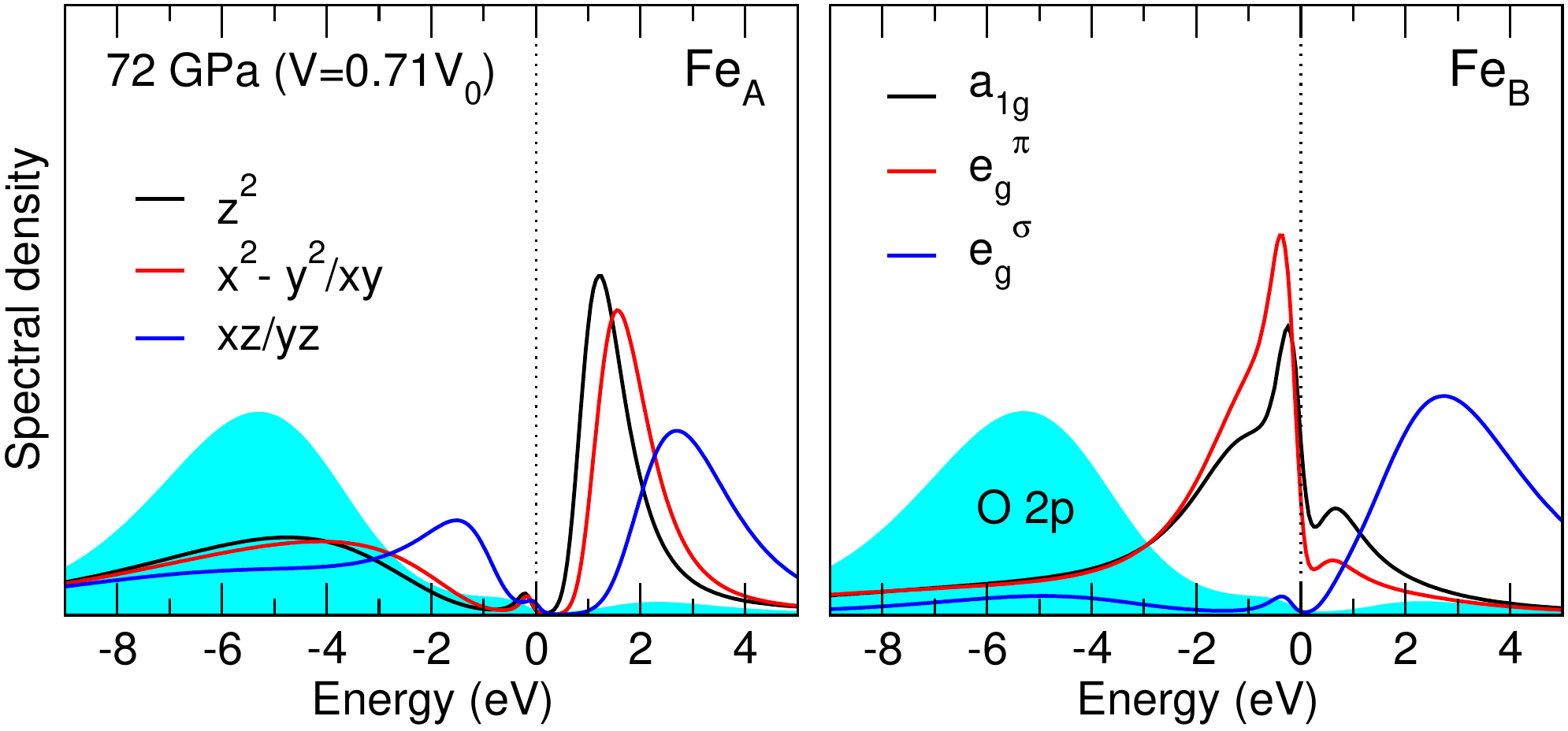}
\caption{Orbitally-resolved spectral functions of paramagnetic DPv Fe$_2$O$_3$ calculated using DFT+DMFT for the Hubbard $U=6$ eV and Hund's rule coupling $J=0.86$ eV at a temperature $T= 1160$ K. The calculated pressure is 72 GPa ($0.71~V_0$). The prismatic Fe $A$ and octahedral Fe $B$ spectral functions are shown by lines. The O $2p$ states are marked in cyan.
}
\label{Fig_S3}
\end{figure}

\begin{figure}[ht]
\centering
\includegraphics[width=1.0\linewidth]{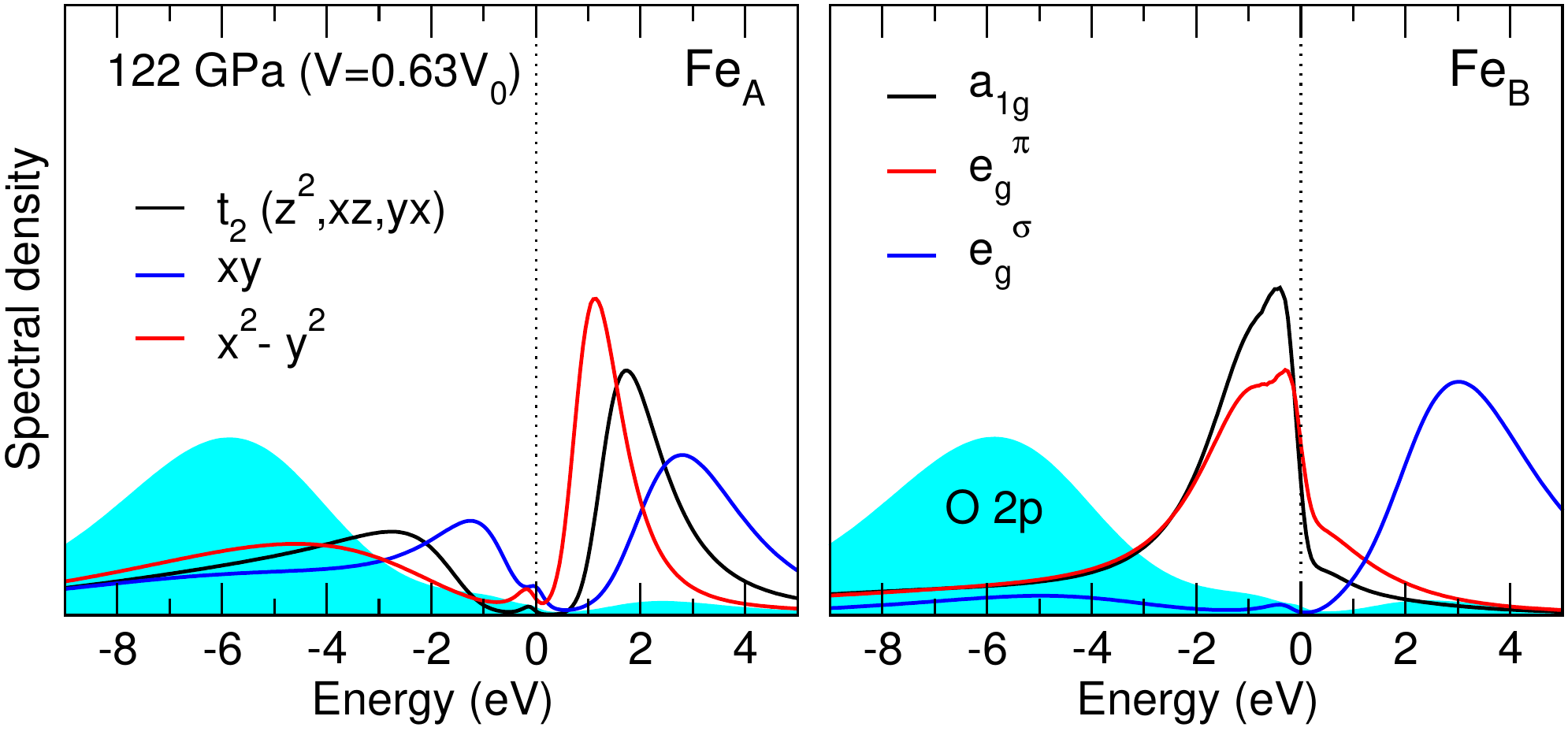}
\caption{Spectral function of paramagnetic PPv Fe$_2$O$_3$ calculated using DFT+DMFT at a temperature $T= 1160$ K ($U=6$ eV and $J=0.86$ eV). The calculated pressure is 122 GPa ($0.63~V_0$). The crystallographically nonequivalent (prismatic) Fe $A$ and octahedral Fe $B$ sites are shown. 
}
\label{Fig_S4}
\end{figure}

We evaluated the crystal field energy splittings for the Fe $3d$ orbitals in the DPv and PPv Fe$_2$O$_3$ as a function of volume. The crystal field splittings are obtained from the first moments of the interacting lattice Green's function for the Fe $3d$ states as $\Delta_{cf} \equiv \mathrm{diag}[ \sum_{\bf k} H^\mathrm{DFT}({\bf k}) + \mathrm{Re} \Sigma( iw_n \rightarrow \infty )]$, where $H({\bf k})$ is the effective low-energy $p$-$d$ Hamiltonian in the Wannier basis set.  $\mathrm{Re} [\Sigma( iw_n \rightarrow \infty )]$ is a static Hartree contribution from self-energy $\Sigma( iw_n )$.
We show our results for the crystal field splittings in Supplementary Fig.~\ref{Fig_S5}. 
Our results for $\Delta_{cf}$ obtained in the non-interacting case, $\Sigma(i\omega_n) \equiv 0$, are shown in Fig.~\ref{Fig_S5} by dotted lines.
In fact, the crystal field splitting is larger for the small-moment Fe $B$ sites with respect to the large-moment Fe $A$ ions. Furthermore, the HS-LS state transition clearly correlates with a remarkable enhancement of the crystal-field splitting by $\sim2$ times (both in the DPv and PPv phases), caused by correlation effects. 
We point out the crucial importance of electronic correlation effects, determined by the contribution of 
$\mathrm{Re}[ \Sigma( iw_n \rightarrow \infty )]$, which play a significant role at the Mott MIT.
If the crystal
field becomes larger than some critical value determined by the Hund's rule coupling $J$, the corresponding Fe-sublattice makes a HS-LS phase transition. Interestingly, our results suggest that the critical value of the crystal filed can approximately be determined from $2 \Delta_{cf} - 10 J = -4 J$ as $\Delta_{cf} = 3 J$ (i.e., from the equality of the energies of the HS and the LS states in the atomic limit).

\begin{figure}[ht]
\centering
\includegraphics[width=1.0\linewidth]{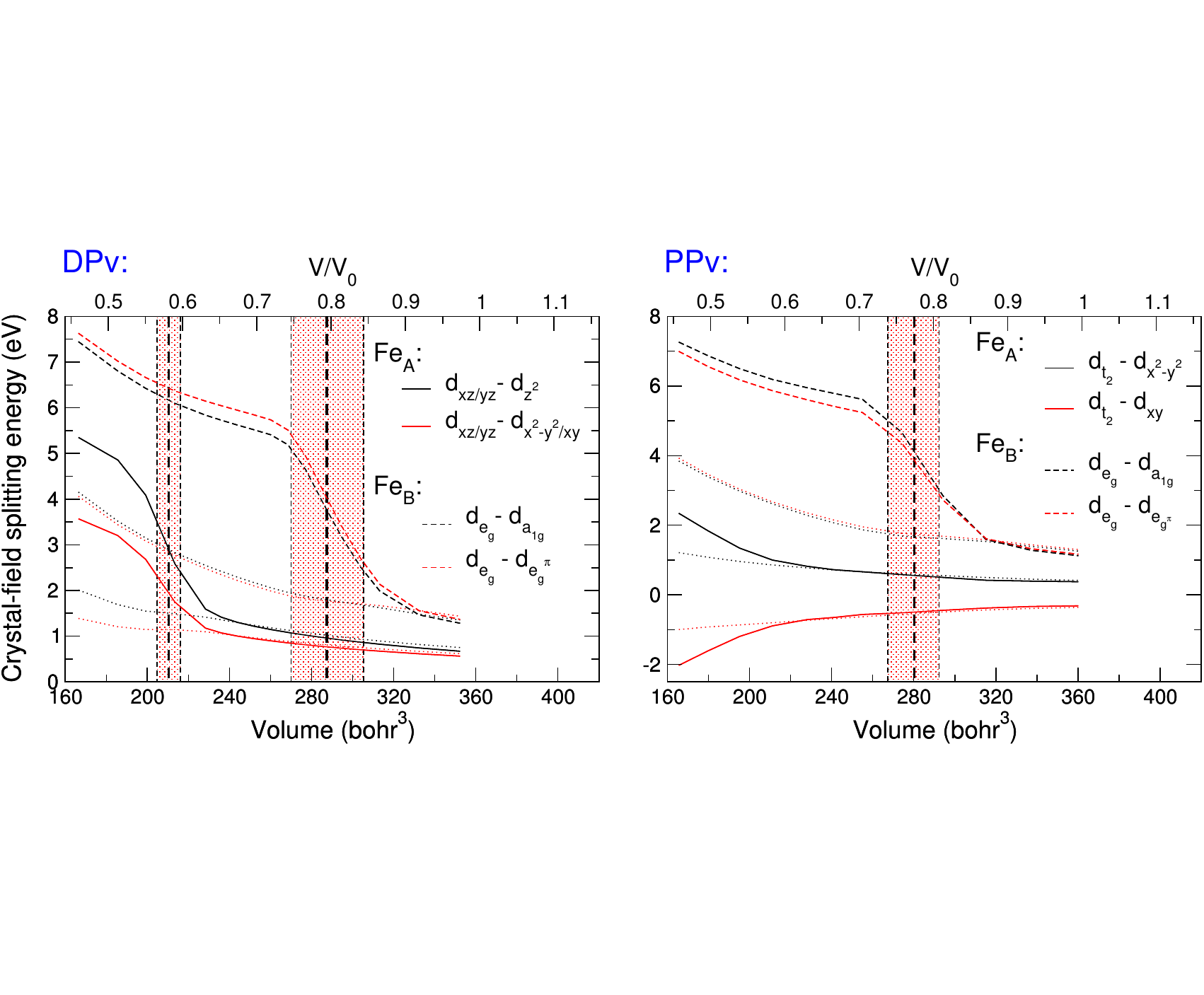}
\caption{Evolution of the effective crystal field splittings of the Wannier Fe $3d$ orbitals obtained by DFT+DMFT for the DPv and PPv phases as a function of volume. The crystal field splittings are determined from the first moments
of the interacting lattice Green's function for the Fe $3d$ states as $\Delta_{cf} \equiv \mathrm{diag}[ \sum_{\bf k} H^\mathrm{DFT}({\bf k}) + \mathrm{Re} \Sigma( iw_n \rightarrow \infty )]$,
where $H({\bf k})$ is the effective low-energy $p$-$d$ Hamiltonian in the Wannier basis set (calculated from the self-consistent DFT+DMFT charge density). We compare our results for $\Delta_{cf}$ with those obtained for the non-interacting case, $\Sigma(i\omega_n) \equiv 0$ (shown by dotted lines). Our results for the lattice volume collapse associated with a site-selective Mott transitions is marked by red shading. 
}
\label{Fig_S5}
\end{figure}

In Supplementary Fig.~\ref{Fig_S6} we show our results for the orbitally-resolved hybridization function [$-\mathrm{Im}(\Delta(\omega))$] for the Fe $A$ and $B$ sites of the DPv (top) and PPv (bottom) phases of Fe$_2$O$_3$.
Our results indicate sufficiently different hybridization strength for the low-lying (e.g., Fe $B$ $t_{2g}$) and the upper-lying (e.g., Fe $B$ $e_g$) orbitals of the Fe $A$ and $B$ sites. This hybridization strength difference is larger for the low-spin Fe $B$ sites with respect to that in the high-spin Fe $A$ site. 
While this difference is sizable near the Fermi level, it becomes even more pronounced near -4 eV$-$ -5 eV below the Fermi level, where the O $2p$ states are located. This sufficiently different hybridization strength seems to mediate the critical value of the crystal field splitting determined from the Hund's coupling $J$ energy scale ($\Delta_{cf} = 3 J$ determined from the equality of the energies of the HS and the LS states in the atomic limit).

\begin{figure}[ht]
\centering
\includegraphics[width=0.9\linewidth]{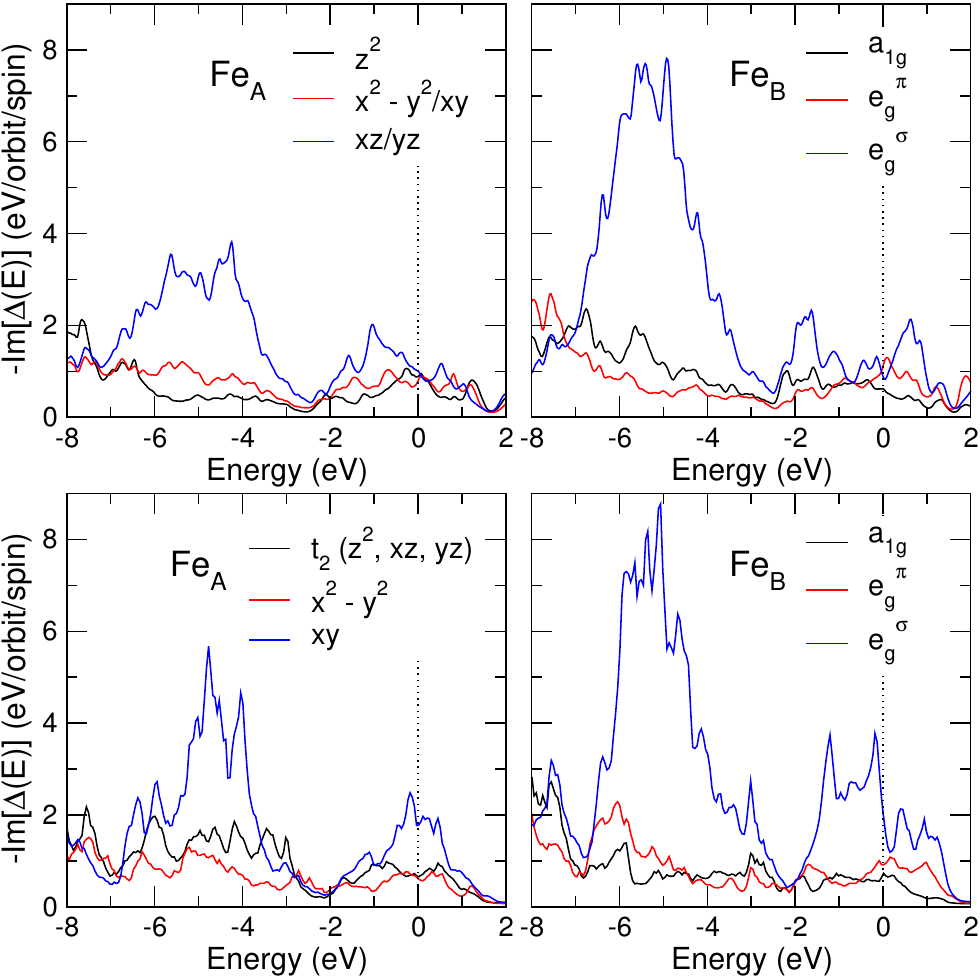}
\caption{
Orbitally-resolved imaginary part of the DMFT hybridization function [-Im($\Delta(\omega)$)] for the Fe $A$ and $B$ sites of the DPv (top) and PPv (bottom) phases of Fe$_2$O$_3$. The DPv results for about 70 GPa and the PPv for $\sim$88 GPa are presented.
}
\label{Fig_S6}
\end{figure}

In our calculations we employ a basis set of atomic-centered symmetry-constrained Wannier functions \cite{RevModPhys.84.1419,PhysRevB.71.125119,JPhysCondMat.20.135227}. For this purpose the localized atomic orbitals of a given symmetry $\phi_{\mu{\bf k}}^\sigma$ (Fe $3d$ and O $2p$) are projected onto the subspace of the Bloch functions $\psi_{i{\bf k}}^\sigma$ (onto the energy bands in a selected energy range near the Fermi level). In this scheme, the Wannier functions are defined as $| w_{\nu{\bf k}}^\sigma \rangle = \sum_{i} P^{\sigma}_{i\nu}(\mathbf{k}) | \psi_{i{\bf k}}^\sigma \rangle$, where $P^{\sigma}_{i\nu}(\mathbf{k})$ are the matrix elements of the projection operator expressed in the basis of local orbitals $\phi^\sigma _{\nu \mathbf{k}} $: $P^{\sigma}_{i\nu}(\mathbf{k}) \equiv \langle \psi_{i{\bf k}}^\sigma | \hat{S}| \phi_{\nu{\bf k}}^\sigma \rangle$ for $i \in [N_1, N_2]$ (for $\varepsilon^\sigma_{i{\bf k}} \in [\varepsilon_1, \varepsilon_2]$) and $P^{\sigma}_{i\nu}(\mathbf{k}) \equiv 0$ otherwise. 

Upon construction of the Wannier basis set we ensure orthogonality of the Wannier orbitals $| w_{\nu{\bf k}}^\sigma \rangle$ by computing the overlap matrix $ O^\sigma_{\mu \nu} (\mathbf{k})= \sum_{i} P^{\sigma*}_{i\mu}(\mathbf{k})  P^{\sigma}_{i\nu}(\mathbf{k})$. This gives us the expression for the orthonormal Wannier functions $| \widetilde{w}_{\nu{\bf k}}^\sigma \rangle$

\begin{eqnarray}
| \widetilde{w}_{\nu{\bf k}}^\sigma \rangle = \sum_{i, \mu} | \psi_{i{\bf k}}^\sigma \rangle P^{\sigma}_{i\mu}(\mathbf{k}) O^{\sigma~-1/2}_{\mu\nu} = \sum_{i} | \psi_{i{\bf k}}^\sigma \rangle \widetilde{P}^{\sigma}_{i\nu}(\mathbf{k}).
\end{eqnarray}

The orthonormal projectors $\widetilde{P}^{\sigma}_{i\nu}(\mathbf{k})$  are used to evaluate the matrix elements of the low-energy DFT Hamiltonian $\hat{H}_{\rm DFT}^\sigma(\mathbf{k})$ within the orthonormal Wannier basis set as
\begin{eqnarray}
\label{eq:hdft}
[ \hat{H}_{\rm DFT}^\sigma(\mathbf{k})]_{\mu \nu} = \sum_{i} \widetilde{P}^{\sigma*}_{i\mu}(\mathbf{k}) \varepsilon^{\sigma}_{i\mathbf{k}} \widetilde{P}^{\sigma}_{i\nu}(\mathbf{k}),
\end{eqnarray}
In our calculations the O $2p$ orbitals were constructed using Wannier functions defined over the full energy range spanned by the $p$-$d$ band complex, whereas the localized Fe $3d$ orbitals are constructed using the Fe $3d$ band set.


\section*{ References}

\begin{thebibliography}{10}
\urlstyle{rm}
\expandafter\ifx\csname url\endcsname\relax
  \def\url#1{\texttt{#1}}\fi
\expandafter\ifx\csname doiprefix\endcsname\relax\def\doiprefix{DOI: }\fi
\providecommand{\bibinfo}[2]{#2}

\bibitem{Nature.436.377}
\bibinfo{author}{Lin, J.-F.} \emph{et~al.}
\newblock \bibinfo{journal}{\bibinfo{title}{{Spin transition of iron in
  magnesiow\" ustite in the Earth's lower mantle}}}.
\newblock {\emph{{Nature}}} \textbf{\bibinfo{volume}{436}},
  \bibinfo{pages}{377}, \doiprefix\url{10.1038/nature03825}
  (\bibinfo{year}{2005}).

\bibitem{GRL.38.L23308}
\bibinfo{author}{Mao, Z.}, \bibinfo{author}{Lin, J.-F.}, \bibinfo{author}{Liu,
  J.} \& \bibinfo{author}{Prakapenka, V.~B.}
\newblock \bibinfo{journal}{\bibinfo{title}{{Thermal equation of state of
  lower-mantle ferropericlase across the spin crossover}}}.
\newblock {\emph{{Geophys. Res. Lett.}}}
  \textbf{\bibinfo{volume}{38}}, \bibinfo{pages}{L23308},
  \doiprefix\url{10.1029/2011GL049915} (\bibinfo{year}{2011}).

\bibitem{PhysRevLett.110.228501}
\bibinfo{author}{Wu, Z.}, \bibinfo{author}{Justo, J.~F.} \&
  \bibinfo{author}{Wentzcovitch, R.~M.}
\newblock \bibinfo{journal}{\bibinfo{title}{{Elastic Anomalies in a
  Spin-Crossover System: Ferropericlase at Lower Mantle Conditions}}}.
\newblock {\emph{{Phys. Rev. Lett.}}}
  \textbf{\bibinfo{volume}{110}}, \bibinfo{pages}{228501},
  \doiprefix\url{10.1103/PhysRevLett.110.228501} (\bibinfo{year}{2013}).

\bibitem{RevGeophys.51.244}
\bibinfo{author}{Lin, J.-F.}, \bibinfo{author}{Speziale, S.},
  \bibinfo{author}{Mao, Z.} \& \bibinfo{author}{Marquardt, H.}
\newblock \bibinfo{journal}{\bibinfo{title}{{Effects of the electronic spin
  transitions of iron in lower mantle minerals: Implications for deep mantle
  geophysics and geochemistry}}}.
\newblock {\emph{{Rev. Geophys.}}} \textbf{\bibinfo{volume}{51}},
  \bibinfo{pages}{244--275}, \doiprefix\url{10.1002/rog.20010}
  (\bibinfo{year}{2013}).

\bibitem{Buffett2007}
\bibinfo{author}{Buffett, B.~A.}
\newblock \bibinfo{journal}{\bibinfo{title}{Earth{\textquoteright}s core and
  the geodynamo}}.
\newblock {\emph{{Science}}} \textbf{\bibinfo{volume}{288}},
  \bibinfo{pages}{2007--2012}, \doiprefix\url{10.1126/science.288.5473.2007}
  (\bibinfo{year}{2000}).


\bibitem{PNAS.106.5508}
\bibinfo{author}{Shim, S.-H.} \emph{et~al.}
\newblock \bibinfo{journal}{\bibinfo{title}{{Electronic and magnetic structures
  of the postperovskite-type Fe$_2$O$_3$ and implications for planetary
  magnetic records and deep interiors}}}.
\newblock {\emph{{Proc. Natl. Acad. Sci. USA}}}
  \textbf{\bibinfo{volume}{106}}, \bibinfo{pages}{5508--5512},
  \doiprefix\url{10.1073/pnas.0808549106} (\bibinfo{year}{2009}).

\bibitem{NatCommun.7.10661}
\bibinfo{author}{Bykova, E.} \emph{et~al.}
\newblock \bibinfo{journal}{\bibinfo{title}{{Structural complexity of simple
  Fe$_2$O$_3$ at high pressures and temperatures}}}.
\newblock {\emph{{Nat. Commun.}}} \textbf{\bibinfo{volume}{7}},
  \bibinfo{pages}{10661}, \doiprefix\url{10.1038/ncomms10661}
  (\bibinfo{year}{2016}).

\bibitem{Nature.570.102}
\bibinfo{author}{Kupenko, I.} \emph{et~al.}
\newblock \bibinfo{journal}{\bibinfo{title}{{Magnetism in cold subducting slabs at mantle transition zone depths}}}.
\newblock {\emph{{Nature}}} \textbf{\bibinfo{volume}{570}},
  \bibinfo{pages}{102--106}, \doiprefix\url{10.1038/s41586-019-1254-8}
  (\bibinfo{year}{2019}).


\bibitem{RevModPhys.70.1039}
\bibinfo{author}{Imada, M.}, \bibinfo{author}{Fujimori, A.} \&
  \bibinfo{author}{Tokura, Y.}
\newblock \bibinfo{journal}{\bibinfo{title}{{Metal-insulator transitions}}}.
\newblock {\emph{{Rev. Mod. Phys.}}}
  \textbf{\bibinfo{volume}{70}}, \bibinfo{pages}{1039--1263},
  \doiprefix\url{10.1103/RevModPhys.70.1039} (\bibinfo{year}{1998}).

\bibitem{PhysRevLett.82.4663}
\bibinfo{author}{Pasternak, M.~P.} \emph{et~al.}
\newblock \bibinfo{journal}{\bibinfo{title}{{Breakdown of the Mott-Hubbard
  State in Fe$_{2}$O$_{3}$: A First-Order Insulator-Metal Transition with
  Collapse of Magnetism at 50 GPa}}}.
\newblock {\emph{{Phys. Rev. Lett.}}}
  \textbf{\bibinfo{volume}{82}}, \bibinfo{pages}{4663--4666},
  \doiprefix\url{10.1103/PhysRevLett.82.4663} (\bibinfo{year}{1999}).

\bibitem{PhysRevLett.102.146402}
\bibinfo{author}{{Kune\v{s}}, J.}, \bibinfo{author}{Korotin, D.~M.},
  \bibinfo{author}{Korotin, M.~A.}, \bibinfo{author}{Anisimov, V.~I.} \&
  \bibinfo{author}{Werner, P.}
\newblock \bibinfo{journal}{\bibinfo{title}{{Pressure-Driven Metal-Insulator
  Transition in Hematite from Dynamical Mean-Field Theory}}}.
\newblock {\emph{{Phys. Rev. Lett.}}}
  \textbf{\bibinfo{volume}{102}}, \bibinfo{pages}{146402},
  \doiprefix\url{10.1103/PhysRevLett.102.146402} (\bibinfo{year}{2009}).

\bibitem{PhysRevX.8.031059}
\bibinfo{author}{Greenberg, E.} \emph{et~al.}
\newblock \bibinfo{journal}{\bibinfo{title}{{Pressure-Induced Site-Selective
  Mott Insulator-Metal Transition in Fe$_{2}$O$_{3}$}}}.
\newblock {\emph{{Phys. Rev. X}}} \textbf{\bibinfo{volume}{8}},
  \bibinfo{pages}{031059}, \doiprefix\url{10.1103/PhysRevX.8.031059}
  (\bibinfo{year}{2018}).

\bibitem{Science.275.654}
\bibinfo{author}{Cohen, R.~E.}, \bibinfo{author}{Mazin, I.~I.} \&
  \bibinfo{author}{Isaak, D.~G.}
\newblock \bibinfo{journal}{\bibinfo{title}{{Magnetic Collapse in Transition
  Metal Oxides at High Pressure: Implications for the Earth}}}.
\newblock {\emph{{Science}}} \textbf{\bibinfo{volume}{275}},
  \bibinfo{pages}{654--657}, \doiprefix\url{10.1126/science.275.5300.654}
  (\bibinfo{year}{1997}).

\bibitem{PhysRevB.92.085142}
\bibinfo{author}{Leonov, I.}
\newblock \bibinfo{journal}{\bibinfo{title}{{Metal-insulator transition and
  local-moment collapse in FeO under pressure}}}.
\newblock {\emph{{Phys. Rev. B}}} \textbf{\bibinfo{volume}{92}},
  \bibinfo{pages}{085142}, \doiprefix\url{10.1103/PhysRevB.92.085142}
  (\bibinfo{year}{2015}).

\bibitem{PhysRevB.94.155135}
\bibinfo{author}{Leonov, I.}, \bibinfo{author}{Pourovskii, L.},
  \bibinfo{author}{Georges, A.} \& \bibinfo{author}{Abrikosov, I.~A.}
\newblock \bibinfo{journal}{\bibinfo{title}{{Magnetic collapse and the behavior
  of transition metal oxides at high pressure}}}.
\newblock {\emph{{Phys. Rev. B}}} \textbf{\bibinfo{volume}{94}},
  \bibinfo{pages}{155135}, \doiprefix\url{10.1103/PhysRevB.94.155135}
  (\bibinfo{year}{2016}).

\bibitem{PhysRevB.96.075136}
\bibinfo{author}{Leonov, I.}, \bibinfo{author}{Ponomareva, A.~V.},
  \bibinfo{author}{Nazarov, R.} \& \bibinfo{author}{Abrikosov, I.~A.}
\newblock \bibinfo{journal}{\bibinfo{title}{{Pressure-induced spin-state
  transition of iron in magnesiow{\"u}stite (Fe,Mg)O}}}.
\newblock {\emph{{Phys. Rev. B}}} \textbf{\bibinfo{volume}{96}},
  \bibinfo{pages}{075136}, \doiprefix\url{10.1103/PhysRevB.96.075136}
  (\bibinfo{year}{2017}).

\bibitem{Jackson_book_1998}
\bibinfo{author}{Jackson, I.}
\newblock \bibinfo{journal}{\bibinfo{title}{Earth's mantle: Composition,
  structure and evolution}}.
\newblock {\emph{{Cambridge University Press}}}
  (\bibinfo{year}{1998}).

\bibitem{PhysRev.83.333}
\bibinfo{author}{Shull, C.~G.}, \bibinfo{author}{Strauser, W.~A.} \&
  \bibinfo{author}{Wollan, E.~O.}
\newblock \bibinfo{journal}{\bibinfo{title}{{Neutron Diffraction by
  Paramagnetic and Antiferromagnetic Substances}}}.
\newblock {\emph{{Phys. Rev.}}} \textbf{\bibinfo{volume}{83}},
  \bibinfo{pages}{333--345}, \doiprefix\url{10.1103/PhysRev.83.333}
  (\bibinfo{year}{1951}).

\bibitem{Greedon_book}
\bibinfo{author}{Greedon, J.~E.}
\newblock \bibinfo{journal}{\bibinfo{title}{{Encyclopedia of Inorganic
  chemistry}}}.
\newblock {\emph{{New York: John Wiley \& Sons}}}
  (\bibinfo{year}{1994}).

\bibitem{PhysRevB.34.7318}
\bibinfo{author}{Fujimori, A.}, \bibinfo{author}{Saeki, M.},
  \bibinfo{author}{Kimizuka, N.}, \bibinfo{author}{Taniguchi, M.} \&
  \bibinfo{author}{Suga, S.}
\newblock \bibinfo{journal}{\bibinfo{title}{{Photoemission satellites and
  electronic structure of Fe$_{2}$O$_{3}$}}}.
\newblock {\emph{{Phys. Rev. B}}} \textbf{\bibinfo{volume}{34}},
  \bibinfo{pages}{7318--7328}, \doiprefix\url{10.1103/PhysRevB.34.7318}
  (\bibinfo{year}{1986}).

\bibitem{PhysRevB.39.13478}
\bibinfo{author}{Lad, R.~J.} \& \bibinfo{author}{Henrich, V.~E.}
\newblock \bibinfo{journal}{\bibinfo{title}{{Photoemission study of the
  valence-band electronic structure in Fe$_{x}$O, Fe$_{3}$O$_{4}$, and
  $\alpha$-Fe$_{2}$O$_{3}$ single crystals}}}.
\newblock {\emph{{Phys. Rev. B}}} \textbf{\bibinfo{volume}{39}},
  \bibinfo{pages}{13478--13485}, \doiprefix\url{10.1103/PhysRevB.39.13478}
  (\bibinfo{year}{1989}).

\bibitem{PhysRevB.66.085115}
\bibinfo{author}{Kim, C.-Y.}, \bibinfo{author}{Bedzyk, M.~J.},
  \bibinfo{author}{Nelson, E.~J.}, \bibinfo{author}{Woicik, J.~C.} \&
  \bibinfo{author}{Berman, L.~E.}
\newblock \bibinfo{journal}{\bibinfo{title}{{Site-specific valence-band
  photoemission study of $\alpha$-Fe$_{2}$O$_{3}$}}}.
\newblock {\emph{{Phys. Rev. B}}} \textbf{\bibinfo{volume}{66}},
  \bibinfo{pages}{085115}, \doiprefix\url{10.1103/PhysRevB.66.085115}
  (\bibinfo{year}{2002}).

\bibitem{PhysRevB.79.035108}
\bibinfo{author}{Gilbert, B.}, \bibinfo{author}{Frandsen, C.},
  \bibinfo{author}{Maxey, E.~R.} \& \bibinfo{author}{Sherman, D.~M.}
\newblock \bibinfo{journal}{\bibinfo{title}{{Band-gap measurements of bulk and
  nanoscale hematite by soft x-ray spectroscopy}}}.
\newblock {\emph{{Phys. Rev. B}}} \textbf{\bibinfo{volume}{79}},
  \bibinfo{pages}{035108}, \doiprefix\url{10.1103/PhysRevB.79.035108}
  (\bibinfo{year}{2009}).

\bibitem{PhysScr.43.327}
\bibinfo{author}{Olsen, J.~S.}, \bibinfo{author}{Cousins, C. S.~G.},
  \bibinfo{author}{Gerward, L.}, \bibinfo{author}{Jhans, H.} \&
  \bibinfo{author}{Sheldon, B.~J.}
\newblock \bibinfo{journal}{\bibinfo{title}{{A study of the crystal structure
  of Fe$_2$O$_3$ in the pressure range up to 65~GPa using synchrotron
  radiation}}}.
\newblock {\emph{{Physica Scripta}}}
  \textbf{\bibinfo{volume}{43}}, \bibinfo{pages}{327},
  \doiprefix\url{10.1088/0031-8949/43/3/021} (\bibinfo{year}{1991}).

\bibitem{PhysRevB.65.064112}
\bibinfo{author}{Rozenberg, G.~K.} \emph{et~al.}
\newblock \bibinfo{journal}{\bibinfo{title}{{High-pressure structural studies
  of hematite Fe$_{2}$O$_{3}$}}}.
\newblock {\emph{{Phys. Rev. B}}} \textbf{\bibinfo{volume}{65}},
  \bibinfo{pages}{064112}, \doiprefix\url{10.1103/PhysRevB.65.064112}
  (\bibinfo{year}{2002}).

\bibitem{PhysRevLett.89.205504}
\bibinfo{author}{Badro, J.} \emph{et~al.}
\newblock \bibinfo{journal}{\bibinfo{title}{{Nature of the High-Pressure
  Transition in Fe$_2$O$_3$ Hematite}}}.
\newblock {\emph{{Phys. Rev. Lett.}}}
  \textbf{\bibinfo{volume}{89}}, \bibinfo{pages}{205504},
  \doiprefix\url{10.1103/PhysRevLett.89.205504} (\bibinfo{year}{2002}).

\bibitem{JPCM.17.269}
\bibinfo{author}{Ono, S.}, \bibinfo{author}{Funakoshi, K.},
  \bibinfo{author}{Ohishi, Y.} \& \bibinfo{author}{Takahashi, E.}
\newblock \bibinfo{journal}{\bibinfo{title}{{In situ x-ray observation of the
  phase transformation of Fe$_2$O$_3$}}}.
\newblock {\emph{{J. Phys.: Condens. Matter}}}
  \textbf{\bibinfo{volume}{17}}, \bibinfo{pages}{269},
  \doiprefix\url{10.1088/0953-8984/17/2/003} (\bibinfo{year}{2005}).

\bibitem{JPCS.66.1714}
\bibinfo{author}{Ono, S.} \& \bibinfo{author}{Ohishi, Y.}
\newblock \bibinfo{journal}{\bibinfo{title}{{In situ X-ray observation of phase
  transformation in Fe$_2$O$_3$ at high pressures and high temperatures}}}.
\newblock {\emph{{J. Phys. Chem. Solids}}}
  \textbf{\bibinfo{volume}{66}}, \bibinfo{pages}{1714--1720},
  \doiprefix\url{10.1016/j.jpcs.2005.06.010} (\bibinfo{year}{2005}).

\bibitem{AmMineral.94.205}
\bibinfo{author}{Ito, E.} \emph{et~al.}
\newblock \bibinfo{journal}{\bibinfo{title}{{Determination of high-pressure
  phase equilibria of Fe$_2$O$_3$ using the Kawai-type apparatus equipped with
  sintered diamond anvils}}}.
\newblock {\emph{{Am. Mineral.}}} \textbf{\bibinfo{volume}{94}},
  \bibinfo{pages}{205--209}, \doiprefix\url{10.2138/am.2009.2913}
  (\bibinfo{year}{2009}).

\bibitem{PhysRevB.94.014112}
\bibinfo{author}{Sanson, A.} \emph{et~al.}
\newblock \bibinfo{journal}{\bibinfo{title}{{Local structure and spin
  transition in Fe$_{2}$O$_{3}$ hematite at high pressure}}}.
\newblock {\emph{{Phys. Rev. B}}} \textbf{\bibinfo{volume}{94}},
  \bibinfo{pages}{014112}, \doiprefix\url{10.1103/PhysRevB.94.014112}
  (\bibinfo{year}{2016}).

\bibitem{JHPR.33.534}
\bibinfo{author}{Bykova, E.} \emph{et~al.}
\newblock \bibinfo{journal}{\bibinfo{title}{{Novel high pressure monoclinic
  Fe$_2$O$_3$ polymorph revealed by single-crystal synchrotron X-ray
  diffraction studies}}}.
\newblock {\emph{{High Press. Res.}}}
  \textbf{\bibinfo{volume}{33}}, \bibinfo{pages}{534--545},
  \doiprefix\url{10.1080/08957959.2013.833613} (\bibinfo{year}{2013}).

\bibitem{SciRep.5.15091}
\bibinfo{author}{Tu\v{c}ek, J.} \emph{et~al.}
\newblock \bibinfo{journal}{\bibinfo{title}{{Zeta-Fe$_2$O$_3$ -- A new stable
  polymorph in iron(III) oxide family}}}.
\newblock {\emph{{Sci. Rep.}}} \textbf{\bibinfo{volume}{5}},
  \bibinfo{pages}{15091}, \doiprefix\url{10.1038/srep15091}
  (\bibinfo{year}{2015}).

\bibitem{PhysRevB.44.943}
\bibinfo{author}{Anisimov, V.~I.}, \bibinfo{author}{Zaanen, J.} \&
  \bibinfo{author}{Andersen, O.~K.}
\newblock \bibinfo{journal}{\bibinfo{title}{{Band theory and Mott insulators:
  Hubbard \textit{U} instead of Stoner \textit{I}}}}.
\newblock {\emph{{Phys. Rev. B}}} \textbf{\bibinfo{volume}{44}},
  \bibinfo{pages}{943--954}, \doiprefix\url{10.1103/PhysRevB.44.943}
  (\bibinfo{year}{1991}).

\bibitem{PhysRevB.69.165107}
\bibinfo{author}{Rollmann, G.}, \bibinfo{author}{Rohrbach, A.},
  \bibinfo{author}{Entel, P.} \& \bibinfo{author}{Hafner, J.}
\newblock \bibinfo{journal}{\bibinfo{title}{{First-principles calculation of
  the structure and magnetic phases of hematite}}}.
\newblock {\emph{{Phys. Rev. B}}} \textbf{\bibinfo{volume}{69}},
  \bibinfo{pages}{165107}, \doiprefix\url{10.1103/PhysRevB.69.165107}
  (\bibinfo{year}{2004}).

\bibitem{PhysRevLett.62.324}
\bibinfo{author}{Metzner, W.} \& \bibinfo{author}{Vollhardt, D.}
\newblock \bibinfo{journal}{\bibinfo{title}{{Correlated Lattice Fermions in
  $d=\ensuremath{\infty}$ Dimensions}}}.
\newblock {\emph{{Phys. Rev. Lett.}}}
  \textbf{\bibinfo{volume}{62}}, \bibinfo{pages}{324--327},
  \doiprefix\url{10.1103/PhysRevLett.62.324} (\bibinfo{year}{1989}).

\bibitem{RevModPhys.68.13}
\bibinfo{author}{Georges, A.}, \bibinfo{author}{Kotliar, G.},
  \bibinfo{author}{Krauth, W.} \& \bibinfo{author}{Rozenberg, M.~J.}
\newblock \bibinfo{journal}{\bibinfo{title}{{Dynamical mean-field theory of
  strongly correlated fermion systems and the limit of infinite dimensions}}}.
\newblock {\emph{{Rev. Mod. Phys.}}}
  \textbf{\bibinfo{volume}{68}}, \bibinfo{pages}{13--125},
  \doiprefix\url{10.1103/RevModPhys.68.13} (\bibinfo{year}{1996}).

\bibitem{RevModPhys.78.865}
\bibinfo{author}{Kotliar, G.} \emph{et~al.}
\newblock \bibinfo{journal}{\bibinfo{title}{{Electronic structure calculations
  with dynamical mean-field theory}}}.
\newblock {\emph{{Rev. Mod. Phys.}}}
  \textbf{\bibinfo{volume}{78}}, \bibinfo{pages}{865--951},
  \doiprefix\url{10.1103/RevModPhys.78.865} (\bibinfo{year}{2006}).

\bibitem{JPCM.9.7359}
\bibinfo{author}{Anisimov, V.~I.}, \bibinfo{author}{Poteryaev, A.~I.},
  \bibinfo{author}{Korotin, M.~A.}, \bibinfo{author}{Anokhin, A.~O.} \&
  \bibinfo{author}{Kotliar, G.}
\newblock \bibinfo{journal}{\bibinfo{title}{{First-principles calculations of
  the electronic structure and spectra of strongly correlated systems:
  dynamical mean-field theory}}}.
\newblock {\emph{{J. Phys.: Condens. Matter}}}
  \textbf{\bibinfo{volume}{9}}, \bibinfo{pages}{7359},
  \doiprefix\url{10.1088/0953-8984/9/35/010} (\bibinfo{year}{1997}).

\bibitem{PhysRevB.57.6884}
\bibinfo{author}{Lichtenstein, A.~I.} \& \bibinfo{author}{Katsnelson, M.~I.}
\newblock \bibinfo{journal}{\bibinfo{title}{{\emph{Ab initio} calculations of
  quasiparticle band structure in correlated systems: LDA++ approach}}}.
\newblock {\emph{{Phys. Rev. B}}} \textbf{\bibinfo{volume}{57}},
  \bibinfo{pages}{6884--6895}, \doiprefix\url{10.1103/PhysRevB.57.6884}
  (\bibinfo{year}{1998}).

\bibitem{EPJST.180.5}
\bibinfo{author}{Kune{\v{s}}, J.} \emph{et~al.}
\newblock \bibinfo{journal}{\bibinfo{title}{{Dynamical mean-field approach to
  materials with strong electronic correlations}}}.
\newblock {\emph{{Eur. Phys. J. Special Topics}}}
  \textbf{\bibinfo{volume}{180}}, \bibinfo{pages}{5--28},
  \doiprefix\url{10.1140/epjst/e2010-01209-0} (\bibinfo{year}{2009}).

\bibitem{PhysRevLett.115.106402}
\bibinfo{author}{Leonov, I.}, \bibinfo{author}{Skornyakov, S.~L.}, \bibinfo{author}{Anisimov, V.~I.} \&
  \bibinfo{author}{Vollhardt, D.}
\newblock \bibinfo{journal}{\bibinfo{title}{{Correlation-Driven Topological Fermi Surface Transition in FeSe}}}.
\newblock {\emph{{Phys. Rev. Lett.}}} \textbf{\bibinfo{volume}{115}},
  \bibinfo{pages}{106402}, \doiprefix\url{10.1103/PhysRevLett.115.106402}
  (\bibinfo{year}{2015}).

\bibitem{PhysRevB.76.235101}
\bibinfo{author}{Pourovskii, L.~V.}, \bibinfo{author}{Amadon, B.},
  \bibinfo{author}{Biermann, S.} \& \bibinfo{author}{Georges, A.}
\newblock \bibinfo{journal}{\bibinfo{title}{{Self-consistency over the charge
  density in dynamical mean-field theory: A linear muffin-tin implementation
  and some physical implications}}}.
\newblock {\emph{{Phys. Rev. B}}} \textbf{\bibinfo{volume}{76}},
  \bibinfo{pages}{235101}, \doiprefix\url{10.1103/PhysRevB.76.235101}
  (\bibinfo{year}{2007}).

\bibitem{PhysRevB.75.155113}
\bibinfo{author}{Haule, K.}
\newblock \bibinfo{journal}{\bibinfo{title}{{Quantum Monte Carlo impurity
  solver for cluster dynamical mean-field theory and electronic structure
  calculations with adjustable cluster base}}}.
\newblock {\emph{{Phys. Rev. B}}} \textbf{\bibinfo{volume}{75}},
  \bibinfo{pages}{155113}, \doiprefix\url{10.1103/PhysRevB.75.155113}
  (\bibinfo{year}{2007}).

\bibitem{PhysRevB.77.205112}
\bibinfo{author}{Amadon, B.} \emph{et~al.}
\newblock \bibinfo{journal}{\bibinfo{title}{{Plane-wave based electronic
  structure calculations for correlated materials using dynamical mean-field
  theory and projected local orbitals}}}.
\newblock {\emph{{Phys. Rev. B}}} \textbf{\bibinfo{volume}{77}},
  \bibinfo{pages}{205112}, \doiprefix\url{10.1103/PhysRevB.77.205112}
  (\bibinfo{year}{2008}).

\bibitem{PhysRevB.80.085101}
\bibinfo{author}{Aichhorn, M.} \emph{et~al.}
\newblock \bibinfo{journal}{\bibinfo{title}{{Dynamical mean-field theory within
  an augmented plane-wave framework: Assessing electronic correlations in the
  iron pnictide LaFeAsO}}}.
\newblock {\emph{{Phys. Rev. B}}} \textbf{\bibinfo{volume}{80}},
  \bibinfo{pages}{085101}, \doiprefix\url{10.1103/PhysRevB.80.085101}
  (\bibinfo{year}{2009}).

\bibitem{PhysRevB.91.195115}
\bibinfo{author}{Leonov, I.}, \bibinfo{author}{Anisimov, V.~I.} \&
  \bibinfo{author}{Vollhardt, D.}
\newblock \bibinfo{journal}{\bibinfo{title}{{Metal-insulator transition and
  lattice instability of paramagnetic V$_{2}$O$_{3}$}}}.
\newblock {\emph{{Phys. Rev. B}}} \textbf{\bibinfo{volume}{91}},
  \bibinfo{pages}{195115}, \doiprefix\url{10.1103/PhysRevB.91.195115}
  (\bibinfo{year}{2015}).

\bibitem{NatCommun.9.4789}
\bibinfo{author}{Bykova, E.} \emph{et~al.}
\newblock \bibinfo{journal}{\bibinfo{title}{{Metastable silica high pressure
  polymorphs as structural proxies of deep Earth silicate melts}}}.
\newblock {\emph{{Nat. Commun.}}} \textbf{\bibinfo{volume}{9}},
  \bibinfo{pages}{4789}, \doiprefix\url{10.1038/s41467-018-07265-z}
  (\bibinfo{year}{2018}).

\bibitem{PhysRevB.66.214422}
\bibinfo{author}{Wright, J.~P.}, \bibinfo{author}{Attfield, J.~P.} \&
  \bibinfo{author}{Radaelli, P.~G.}
\newblock \bibinfo{journal}{\bibinfo{title}{{Charge ordered structure of
  magnetite Fe$_{3}$O$_{4}$ below the Verwey transition}}}.
\newblock {\emph{{Phys. Rev. B}}} \textbf{\bibinfo{volume}{66}},
  \bibinfo{pages}{214422}, \doiprefix\url{10.1103/PhysRevB.66.214422}
  (\bibinfo{year}{2002}).

\bibitem{Nature.481.173}
\bibinfo{author}{Senn, M.~S.},\bibinfo{author}{Wright, J.~P.} \&
 \bibinfo{author}{Attfield, J.~P.} 
\newblock \bibinfo{journal}{\bibinfo{title}{{Charge order and three-site distortions in the Verwey structure of magnetite}}}.
\newblock {\emph{{Nature}}} \textbf{\bibinfo{volume}{481}},
  \bibinfo{pages}{173}, \doiprefix\url{10.1038/nature10704}
  (\bibinfo{year}{2012}).


\bibitem{NatCommun.10.2857}
\bibinfo{author}{Perversi, G.} \emph{et~al.}
\newblock \bibinfo{journal}{\bibinfo{title}{{Co-emergence of magnetic order and structural fluctuations in magnetite}}}.
\newblock {\emph{{Nat. Commun.}}} \textbf{\bibinfo{volume}{10}},
  \bibinfo{pages}{2857}, \doiprefix\url{10.1038/s41467-019-10949-9}
  (\bibinfo{year}{2019}).


\bibitem{1367-2630-7-1-053}
\bibinfo{author}{Radaelli, P.~G.}
\newblock \bibinfo{journal}{\bibinfo{title}{Orbital ordering in
  transition-metal spinels}}.
\newblock {\emph{{New J. Phys.}}}
  \textbf{\bibinfo{volume}{7}}, \bibinfo{pages}{53},
  \doiprefix\url{10.1088/1367-2630/7/1/053} (\bibinfo{year}{2005}).

\bibitem{PhysRevB.74.165117}
\bibinfo{author}{Leonov, I.}, \bibinfo{author}{Yaresko, A.~N.},
  \bibinfo{author}{Antonov, V.~N.} \& \bibinfo{author}{Anisimov, V.~I.}
\newblock \bibinfo{journal}{\bibinfo{title}{{Electronic structure of
  charge-ordered Fe$_{3}$O$_{4}$ from calculated optical, magneto-optical Kerr
  effect, and O $K$-edge x-ray absorption spectra}}}.
\newblock {\emph{{Phys. Rev. B}}} \textbf{\bibinfo{volume}{74}},
  \bibinfo{pages}{165117}, \doiprefix\url{10.1103/PhysRevB.74.165117}
  (\bibinfo{year}{2006}).

\bibitem{PhysRevB.74.195115}
\bibinfo{author}{Jeng, H.-T.}, \bibinfo{author}{Guo, G.~Y.},
  \& \bibinfo{author}{Huang, D.~J.}
\newblock \bibinfo{journal}{\bibinfo{title}{{Charge-orbital ordering in low-temperature structures of magnetite: GGA+$U$ investigations}}}.
\newblock {\emph{{Phys. Rev. B}}} \textbf{\bibinfo{volume}{74}},
  \bibinfo{pages}{195115}, \doiprefix\url{10.1103/PhysRevB.74.195115}
  (\bibinfo{year}{2006}).


\bibitem{doi:10.1063/1.3584855}
\bibinfo{author}{Pontius, N.} \emph{et~al.}
\newblock \bibinfo{journal}{\bibinfo{title}{{Time-resolved resonant soft x-ray
  diffraction with free-electron lasers: Femtosecond dynamics across the Verwey
  transition in magnetite}}}.
\newblock {\emph{{Appl. Phys. Lett.}}}
  \textbf{\bibinfo{volume}{98}}, \bibinfo{pages}{182504},
  \doiprefix\url{10.1063/1.3584855} (\bibinfo{year}{2011}).

\bibitem{PhysRevB.72.014407}
\bibinfo{author}{Leonov, I.}, \bibinfo{author}{Yaresko, A.~N.},
  \bibinfo{author}{Antonov, V.~N.}, \bibinfo{author}{Attfield, J.~P.} \&
  \bibinfo{author}{Anisimov, V.~I.}
\newblock \bibinfo{journal}{\bibinfo{title}{{Charge order in Fe$_{2}$OBO$_{3}$:
  An LSDA+$U$ study}}}.
\newblock {\emph{{Phys. Rev. B}}} \textbf{\bibinfo{volume}{72}},
  \bibinfo{pages}{014407}, \doiprefix\url{10.1103/PhysRevB.72.014407}
  (\bibinfo{year}{2005}).

\bibitem{PhysRevLett.106.256401}
\bibinfo{author}{{Kune\v{s}}, J.} \& \bibinfo{author}{{K\v{r}\'apek}, V.}
\newblock \bibinfo{journal}{\bibinfo{title}{{Disproportionation and
  Metallization at Low-Spin to High-Spin Transition in Multiorbital Mott
  Systems}}}.
\newblock {\emph{{Phys. Rev. Lett.}}}
  \textbf{\bibinfo{volume}{106}}, \bibinfo{pages}{256401},
  \doiprefix\url{10.1103/PhysRevLett.106.256401} (\bibinfo{year}{2011}).

\bibitem{PhysRevB.86.205131}
\bibinfo{author}{Ovsyannikov, S.~V.}, \bibinfo{author}{Morozova, N.~V.},
  \bibinfo{author}{Karkin, A.~E.} \& \bibinfo{author}{Shchennikov, V.~V.}
\newblock \bibinfo{journal}{\bibinfo{title}{{High-pressure cycling of hematite
  $\ensuremath{\alpha}$-Fe$_{2}$O$_{3}$: Nanostructuring, in situ electronic
  transport, and possible charge disproportionation}}}.
\newblock {\emph{{Phys. Rev. B}}} \textbf{\bibinfo{volume}{86}},
  \bibinfo{pages}{205131}, \doiprefix\url{10.1103/PhysRevB.86.205131}
  (\bibinfo{year}{2012}).

\bibitem{PhysRevLett.109.156402}
\bibinfo{author}{Park, H.}, \bibinfo{author}{Millis, A.~J.} \&
  \bibinfo{author}{Marianetti, C.~A.}
\newblock \bibinfo{journal}{\bibinfo{title}{Site-selective mott transition in
  rare-earth-element nickelates}}.
\newblock {\emph{{Phys. Rev. Lett.}}}
  \textbf{\bibinfo{volume}{109}}, \bibinfo{pages}{156402},
  \doiprefix\url{10.1103/PhysRevLett.109.156402} (\bibinfo{year}{2012}).

\bibitem{PhysRevB.91.075128}
\bibinfo{author}{Subedi, A.}, \bibinfo{author}{Peil, O.~E.} \&
  \bibinfo{author}{Georges, A.}
\newblock \bibinfo{journal}{\bibinfo{title}{Low-energy description of the
  metal-insulator transition in the rare-earth nickelates}}.
\newblock {\emph{{Phys. Rev. B}}} \textbf{\bibinfo{volume}{91}},
  \bibinfo{pages}{075128}, \doiprefix\url{10.1103/PhysRevB.91.075128}
  (\bibinfo{year}{2015}).

\bibitem{PhysRevB.92.155145}
\bibinfo{author}{Ruppen, J.} \emph{et~al.}
\newblock \bibinfo{journal}{\bibinfo{title}{Optical spectroscopy and the nature
  of the insulating state of rare-earth nickelates}}.
\newblock {\emph{{Phys. Rev. B}}} \textbf{\bibinfo{volume}{92}},
  \bibinfo{pages}{155145}, \doiprefix\url{10.1103/PhysRevB.92.155145}
  (\bibinfo{year}{2015}).

\bibitem{PhysRevB.96.205139}
\bibinfo{author}{Seth, P.} \emph{et~al.}
\newblock \bibinfo{journal}{\bibinfo{title}{Renormalization of effective
  interactions in a negative charge transfer insulator}}.
\newblock {\emph{{Phys. Rev. B}}} \textbf{\bibinfo{volume}{96}},
  \bibinfo{pages}{205139}, \doiprefix\url{10.1103/PhysRevB.96.205139}
  (\bibinfo{year}{2017}).

\bibitem{PhysRevB.96.045120}
\bibinfo{author}{Ruppen, J.} \emph{et~al.}
\newblock \bibinfo{journal}{\bibinfo{title}{Impact of antiferromagnetism on the
  optical properties of rare-earth nickelates}}.
\newblock {\emph{{Phys. Rev. B}}} \textbf{\bibinfo{volume}{96}},
  \bibinfo{pages}{045120}, \doiprefix\url{10.1103/PhysRevB.96.045120}
  (\bibinfo{year}{2017}).

\bibitem{JPhysCondMat.21.395502}
\bibinfo{author}{Giannozzi, P.} \emph{et~al.}
\newblock \bibinfo{journal}{\bibinfo{title}{{QUANTUM ESPRESSO: a modular and
  open-source software project for quantum simulations of materials}}}.
\newblock {\emph{{J. Phys. Condens. Matter}}}
  \textbf{\bibinfo{volume}{21}}, \bibinfo{pages}{395502},
  \doiprefix\url{10.1088/0953-8984/21/39/395502} (\bibinfo{year}{2009}).

\bibitem{PhysRevLett.101.096405}
\bibinfo{author}{Leonov, I.} \emph{et~al.}
\newblock \bibinfo{journal}{\bibinfo{title}{{Structural Relaxation due to
  Electronic Correlations in the Paramagnetic Insulator KCuF$_{3}$}}}.
\newblock {\emph{{Phys. Rev. Lett.}}}
  \textbf{\bibinfo{volume}{101}}, \bibinfo{pages}{096405},
  \doiprefix\url{10.1103/PhysRevLett.101.096405} (\bibinfo{year}{2008}).

\bibitem{PhysRevB.81.075109}
\bibinfo{author}{Leonov, I.}, \bibinfo{author}{Korotin, D.},
  \bibinfo{author}{Binggeli, N.}, \bibinfo{author}{Anisimov, V.~I.} \&
  \bibinfo{author}{Vollhardt, D.}
\newblock \bibinfo{journal}{\bibinfo{title}{{Computation of correlation-induced
  atomic displacements and structural transformations in paramagnetic
  KCuF$_{3}$ and LaMnO$_{3}$}}}.
\newblock {\emph{{Phys. Rev. B}}} \textbf{\bibinfo{volume}{81}},
  \bibinfo{pages}{075109}, \doiprefix\url{10.1103/PhysRevB.81.075109}
  (\bibinfo{year}{2010}).

\bibitem{RevModPhys.84.1419}
\bibinfo{author}{Marzari, N.}, \bibinfo{author}{Mostofi, A.~A.},
  \bibinfo{author}{Yates, J.~R.}, \bibinfo{author}{Souza, I.} \&
  \bibinfo{author}{Vanderbilt, D.}
\newblock \bibinfo{journal}{\bibinfo{title}{{Maximally localized Wannier
  functions: Theory and applications}}}.
\newblock {\emph{{Rev. Mod. Phys.}}}
  \textbf{\bibinfo{volume}{84}}, \bibinfo{pages}{1419--1475},
  \doiprefix\url{10.1103/RevModPhys.84.1419} (\bibinfo{year}{2012}).

\bibitem{PhysRevB.71.125119}
\bibinfo{author}{Anisimov, V.~I.} \emph{et~al.}
\newblock \bibinfo{journal}{\bibinfo{title}{{Full orbital calculation scheme
  for materials with strongly correlated electrons}}}.
\newblock {\emph{{Phys. Rev. B}}} \textbf{\bibinfo{volume}{71}},
  \bibinfo{pages}{125119}, \doiprefix\url{10.1103/PhysRevB.71.125119}
  (\bibinfo{year}{2005}).

\bibitem{JPhysCondMat.20.135227}
\bibinfo{author}{Trimarchi, G.}, \bibinfo{author}{Leonov, I.},
  \bibinfo{author}{Binggeli, N.}, \bibinfo{author}{Korotin, D.} \&
  \bibinfo{author}{Anisimov, V.~I.}
\newblock \bibinfo{journal}{\bibinfo{title}{{LDA+DMFT implemented with the
  pseudopotential plane-wave approach}}}.
\newblock {\emph{{J. Phys. Condens. Matter}}}
  \textbf{\bibinfo{volume}{20}}, \bibinfo{pages}{135227},
  \doiprefix\url{10.1088/0953-8984/20/13/135227} (\bibinfo{year}{2008}).

\bibitem{RevModPhys.83.349}
\bibinfo{author}{Gull, E.} \emph{et~al.}
\newblock \bibinfo{journal}{\bibinfo{title}{{Continuous-time Monte Carlo
  methods for quantum impurity models}}}.
\newblock {\emph{{Rev. Mod. Phys.}}}
  \textbf{\bibinfo{volume}{83}}, \bibinfo{pages}{349--404},
  \doiprefix\url{10.1103/RevModPhys.83.349} (\bibinfo{year}{2011}).

\end{thebibliography}

\begin{thebibliography}{10}
\urlstyle{rm}
\expandafter\ifx\csname url\endcsname\relax
  \def\url#1{\texttt{#1}}\fi
\expandafter\ifx\csname doiprefix\endcsname\relax\def\doiprefix{DOI: }\fi
\providecommand{\bibinfo}[2]{#2}

\bibitem{PhysRevLett.102.146402}
\bibinfo{author}{{Kune\v{s}}, J.}, \bibinfo{author}{Korotin, D.~M.},
  \bibinfo{author}{Korotin, M.~A.}, \bibinfo{author}{Anisimov, V.~I.} \&
  \bibinfo{author}{Werner, P.}
\newblock \bibinfo{journal}{\bibinfo{title}{{Pressure-Driven Metal-Insulator
  Transition in Hematite from Dynamical Mean-Field Theory}}}.
\newblock {\emph{{Phys. Rev. Lett.}}}
  \textbf{\bibinfo{volume}{102}}, \bibinfo{pages}{146402},
  \doiprefix\url{10.1103/PhysRevLett.102.146402} (\bibinfo{year}{2009}).

\bibitem{PhysRevX.8.031059}
\bibinfo{author}{Greenberg, E.} \emph{et~al.}
\newblock \bibinfo{journal}{\bibinfo{title}{{Pressure-Induced Site-Selective
  Mott Insulator-Metal Transition in Fe$_{2}$O$_{3}$}}}.
\newblock {\emph{{Phys. Rev. X}}} \textbf{\bibinfo{volume}{8}},
  \bibinfo{pages}{031059}, \doiprefix\url{10.1103/PhysRevX.8.031059}
  (\bibinfo{year}{2018}).

\bibitem{RevModPhys.84.1419}
\bibinfo{author}{Marzari, N.}, \bibinfo{author}{Mostofi, A.~A.},
  \bibinfo{author}{Yates, J.~R.}, \bibinfo{author}{Souza, I.} \&
  \bibinfo{author}{Vanderbilt, D.}
\newblock \bibinfo{journal}{\bibinfo{title}{{Maximally localized Wannier
  functions: Theory and applications}}}.
\newblock {\emph{{Rev. Mod. Phys.}}}
  \textbf{\bibinfo{volume}{84}}, \bibinfo{pages}{1419--1475},
  \doiprefix\url{10.1103/RevModPhys.84.1419} (\bibinfo{year}{2012}).

\bibitem{PhysRevB.71.125119}
\bibinfo{author}{Anisimov, V.~I.} \emph{et~al.}
\newblock \bibinfo{journal}{\bibinfo{title}{{Full orbital calculation scheme
  for materials with strongly correlated electrons}}}.
\newblock {\emph{{Phys. Rev. B}}} \textbf{\bibinfo{volume}{71}},
  \bibinfo{pages}{125119}, \doiprefix\url{10.1103/PhysRevB.71.125119}
  (\bibinfo{year}{2005}).

\bibitem{JPhysCondMat.20.135227}
\bibinfo{author}{Trimarchi, G.}, \bibinfo{author}{Leonov, I.},
  \bibinfo{author}{Binggeli, N.}, \bibinfo{author}{Korotin, D.} \&
  \bibinfo{author}{Anisimov, V.~I.}
\newblock \bibinfo{journal}{\bibinfo{title}{{LDA+DMFT implemented with the
  pseudopotential plane-wave approach}}}.
\newblock {\emph{{J. Phys. Condens. Matter}}}
  \textbf{\bibinfo{volume}{20}}, \bibinfo{pages}{135227},
  \doiprefix\url{10.1088/0953-8984/20/13/135227} (\bibinfo{year}{2008}).

\end{thebibliography}
\end{document}